\documentclass[11pt,document,nofootinbib,superscriptaddress,onecolumn,preprintnumbers,balancelastpage]{article}
\pdfoutput=1
\usepackage{jheppub}
\usepackage{graphicx}
\usepackage{amsmath} 
\usepackage{perpage}
\usepackage[T1]{fontenc} 
\usepackage{amssymb}
\usepackage{color}
\usepackage{cleveref}
\usepackage{feynmf}
\usepackage{upgreek}
\usepackage{listings}
\usepackage{ctable}
\usepackage{lipsum}
\usepackage{slashed}
\usepackage{mathtools}
\usepackage[caption=false]{subfig}

\newcommand*\colvec[3][]{\begin{pmatrix}\ifx\relax#1\relax\else#1\\\fi#2\\#3\end{pmatrix}}
\newcommand{\beq}{\begin{equation}}
\newcommand{\beqn}{\begin{eqnarray}}
\newcommand{\eeq}{\end{equation}}
\newcommand{\eeqn}{\end{eqnarray}}

\definecolor{darklavender}{rgb}{0.45, 0.31, 0.59}

\newcommand{\Neff}{N_\text{eff}}
\newcommand{\SM}{\text{SM}}
\newcommand{\inv}{\text{inv}}
\newcommand{\dec}{\text{dec}}
\newcommand{\mA}{m_{\gamma '}}

\newcommand{\dd}{\partial}

\newcommand{\Lag}{\mathcal{L}}
\newcommand{\zsym}{\mathbf{Z}_2}
\newcommand{\one}{\text{U}(1)}
\newcommand{\su}{\text{SU}}
\newcommand{\prn}[1]{ \left(  #1 \right) }

\newcommand{\ord}[1]{\mathcal{O}\left(#1 \right)}
\newcommand{\pmat}[1]{\begin{pmatrix} #1 \end{pmatrix}}
\newcommand{\evat}[1]{\left. #1 \right|}

\newcommand{\al}[1]{\begin{align} #1 \end{align}}

\title{A Predictive Mirror Twin Higgs with Small $\mathbf{Z}_2$ Breaking}
\author[1,2,3]{Keisuke Harigaya, }
\author[2,3]{Robert McGehee, }
\author[2,3,4,\ast]{Hitoshi Murayama, }
\author[2,3]{and Katelin Schutz}
\affiliation[1]{School of Natural Sciences, Institute for Advanced Study, Princeton, NJ, 08540}
\affiliation[2]{Berkeley Center for Theoretical Physics, University of California, Berkeley, CA 94720}
\affiliation[3]{Theoretical Physics Group, Lawrence Berkeley National Laboratory, Berkeley, CA 94720}
\affiliation[4]{Kavli Institute for the Physics and Mathematics of the Universe (WPI),
University of Tokyo Institutes for Advanced Study, University of Tokyo,
Kashiwa 277-8583, Japan}
\affiliation[\ast]{Hamamatsu Professor}

\abstract{The twin Higgs mechanism is a solution to the little hierarchy problem in which the top partner is neutral under the Standard Model (SM) gauge group. The simplest mirror twin Higgs (MTH) model -- where a $\zsym$ symmetry copies each SM particle -- has too many relativistic degrees of freedom to be consistent with cosmological observations. We demonstrate that MTH models can have an observationally viable cosmology if the twin mass spectrum leads to twin neutrino decoupling before the SM and twin QCD phase transitions. Our solution requires the twin photon to have a mass of $\sim 20$~MeV and kinetically mix with the SM photon to mediate entropy transfer from the twin sector to the SM. This twin photon can be robustly discovered or excluded by future experiments. Additionally, the residual twin degrees of freedom present in the early Universe in this scenario would be detectable by future observations of the cosmic microwave background.}

\begin{document}
\maketitle
\flushbottom
\section{Introduction}

The disparity between the electroweak scale and the Planck scale is one of the most outstanding problems in particle physics (see, {\it e.g.}\/, \cite{Murayama:2000dw}). Explanations have been provided by both supersymmetry and the compositeness of the Higgs, where the electroweak scale originates from a supersymmetry breaking scale~\cite{Maiani:1979cx,Veltman:1980mj,Witten:1981nf,Kaul:1981wp} or a composite scale~\cite{Kaplan:1983fs,Kaplan:1983sm}. Without fine-tuning parameters, these classes of solutions generically predict the existence of a partner to the top quark that is colored and as light as the electroweak scale. Such a particle has not been observed at the Large Hadron Collider (LHC), providing a strong lower bound on its mass, typically around 1 TeV \cite{Aaboud:2018kya,Sirunyan:2018vjp,Aad:2016shx,Sirunyan:2018omb}. In order to accommodate this bound, these kinds of theories require fine-tuning of their parameters to fix the electroweak scale. The need for this fine-tuning is called the little hierarchy problem.

The twin Higgs mechanism~\cite{Chacko:2005pe} addresses this problem. The mechanism is based on a $\zsym$ symmetry that introduces a copy of the Standard Model (SM) particles which we call twin particles, and an approximate global symmetry of the scalar potential. After the twin Higgs obtains a vacuum expectation value (VEV), $f$, the SM Higgs becomes a pseudo-Nambu-Goldstone boson, protecting the Higgs mass from quantum corrections up to the scale $\Lambda_{\text{TH}} \approx 4\pi f$. The top quark partner is now twin colored and not easily produced at the LHC, thereby solving the little hierarchy problem. The twin Higgs mechanism is readily incorporated into solutions of the full hierarchy problem; for instance, supersymmetric~\cite{Falkowski:2006qq,Chang:2006ra,Craig:2013fga,Katz:2016wtw,Badziak:2017syq,Badziak:2017kjk,Badziak:2017wxn} and composite~\cite{Batra:2008jy,Geller:2014kta,Barbieri:2015lqa,Low:2015nqa,Cheng:2015buv,Csaki:2015gfd,Cheng:2016uqk,Contino:2017moj} realizations of the idea have been explored.

While the twin Higgs mechanism is theoretically appealing, it is difficult to reconcile with cosmological observations. The simplest realization of this scenario is the mirror twin Higgs (MTH) model where the $\zsym$ symmetry is a fundamental symmetry (as opposed to an emergent symmetry).
Twin particles thermalize with SM particles via Higgs exchange in the early Universe. The fundamental $\zsym$ symmetry predicts that the entropy of light twin particles is eventually transferred into twin photons and twin neutrinos, which behave as extra radiation components. During epochs when the Universe is radiation dominated, these extra radiation components contribute appreciably to the expansion of the Universe. The expansion rate depends on the energy density in relativistic species, which is typically parameterized in relation to the photon energy density as 
\beq
\label{eq:Neffdefn}
\rho_r = \left(1 + \frac{7}{8} \left(\frac{4}{11}\right)^{4/3} \Neff\right) \rho_\gamma, 
\eeq 
where $\Neff$ is the effective number of (light) neutrino species, 
the factor of $7/8$ comes from Fermi-Dirac statistics, and the factor of $(4/11)^{4/3}$ comes from the fact that electron-positron pairs annihilate after SM neutrino decoupling and heat the photons. SM neutrinos are still partially in thermal equilibrium with the rest of the thermal bath when the electron-positron pairs start to annihilate, which yields a predicted SM value of $\Neff \approx 3.046$~\cite{deSalas:2016ztq,Mangano:2001iu}. Meanwhile, in the MTH model, the additional number of relativistic species (twin photons and twin neutrinos) modify the SM prediction by an amount $\Delta \Neff \sim 5.6$ \cite{Chacko:2016hvu}. Big bang nucleosynthesis (BBN) and anisotropies in the cosmic microwave background (CMB) are both exquisitely sensitive to the expansion history during epochs when the energy density in radiation was non-negligible, and provide independent measurements of $\Neff$. The BBN measurement of $\Neff$ (including the observed helium and deuterium abundances) is $2.85\pm0.28$~\cite{Cyburt:2015mya}. Meanwhile, the \emph{Planck}~2018 measurement of $\Neff$ (from $TT$, $TE$, and $EE$ power spectra combined with lensing and baryon acoustic oscillations) is $\Neff = 2.99^{+0.34}_{-0.33}$ at 95\% confidence~\cite{planck2018}. Both of these measurements indicate that the MTH scenario is excluded at high significance. 

A number of ways to reduce the twin contribution to $\Neff$ have been explored. For instance, the fraternal twin Higgs (FTH) mechanism~\cite{Craig:2015pha} lacks the first and second generations of twin fermions and also lacks a twin photon. The single twin neutrino yields $\Delta \Neff \approx 0.075$ which is still consistent with observations \cite{Craig:2015xla}. However, given the lack of $\zsym$ symmetry, the proximity of the top Yukawa, $\su (2)$ gauge, and $\su (3)$ gauge couplings in the SM and twin sectors should be addressed. One could also make the twin neutrinos heavy~\cite{Barbieri:2016zxn,Csaki:2017spo} and even the twin photon heavy without affecting naturalness~\cite{Batell:2019ptb,Liu:2019ixm}, while expanding the possibilities for twin dark matter candidates \cite{Hochberg:2018vdo,Cheng:2018vaj}. Asymmetric entropy production after the twin and SM sectors decouple is another way to diminish $\Neff$~\cite{Chacko:2016hvu,Craig:2016lyx,Craig:2016kue,Csaki:2017spo,Koren:2019iuv}. Its effects on the matter power spectrum could also be seen by future large scale structure observations \cite{Chacko:2018vss}. Refs.~\cite{Barbieri:2016zxn,Barbieri:2017opf} investigate the Minimal MTH where the twin Yukawa couplings are raised, which reduces $\Delta \Neff$ because there are few twin degrees of freedom when the SM and twin sectors decouple from each other. The (nearly) massless twin photons and neutrinos still contribute appreciably to $\Neff$, which is at least $3.3$ and in slight tension with the \emph{Planck} measurement.

In this paper, we consider a MTH model with a fundamental $\zsym$ symmetry at high energies which is preserved as much as possible at the electroweak scale. 
As is shown in Ref.~\cite{Barbieri:2016zxn}, it is mandatory to increase the twin Yukawa couplings (except for the twin top) to suppress $\Delta \Neff$. Note that the contribution of twin Yukawa couplings $\lesssim 0.1$ to the Higgs mass squared does not reintroduce fine-tuning below $\Lambda_{\text{TH}}$. To build on the models explored in Refs.~\cite{Barbieri:2016zxn,Barbieri:2017opf}, we give the twin photon a Stueckelberg mass. This allows entropy from the twin QCD phase transition to transfer into the SM via decaying twin photons, thereby minimizing $\Delta \Neff$ while achieving a minimal $\zsym$ breaking.

We keep the $\zsym $ breaking as minimal as possible and do not consider $\zsym$ breaking gauge couplings. This not not only motivated by minimality, but also by the theory of flavor. The hierarchy of the Yukawa couplings in the SM is one of its great mysteries which can be explained by introducing some fields whose values control the Yukawa couplings, like in the Froggatt-Nielsen mechanism~\cite{Froggatt:1978nt}. In such a mechanism, it is possible that the field controlling the SM Yukawa couplings spontaneously takes different values from its twin counterpart which sets the twin Yukawa couplings~\cite{Barbieri:2017opf}. This scheme naturally maintains $y_t \sim y'_t$ necessary for the twin Higgs mechanism. We could also introduce moduli fields whose values control the gauge couplings perhaps motivated by string theory, but these would differ from the Yukawa-setting fields in the Froggatt-Nielsen mechanism in that these moduli would not be motivated by low-energy, known SM problems.

We assume that the twin neutrinos are effectively massless, as is the case in the SM, motivated by the following observation.
Let us assume that the neutrino mass originates from a see-saw mechanism~\cite{Yanagida:1979as,GellMann:1980vs,Minkowski:1977sc,Mohapatra:1979ia}. 
We may raise the twin neutrino masses by smaller twin right-handed neutrino masses. Raising the twin neutrino masses to a level where they do not contribute to $\Neff$ requires significant $\zsym$ breaking right-handed neutrino masses, in contrast to the situation described above, and care must be taken to avoid spoiling the twin Higgs mechanism.
We may instead raise the twin neutrino masses by a larger yukawa coupling of the twin right-handed neutrino to the twin left-handed neutrino and the twin Higgs. Let us consider a well-motivated benchmark point of thermal leptogenesis~\cite{Fukugita:1986hr}, which requires the SM yukawa coupling $y_N$ to satisfy $y_N^2 > 10^{-5}$~\cite{Giudice:2003jh,Buchmuller:2004nz}. Then even if the twin $y_N= O(1)$ and $f/v = 10$, the mirror neutrino mass is at the most $0.1 \text{ eV} \cdot 10^5 \cdot (10)^2 = \text{ MeV}$, which is not large enough to evade cosmological bounds. For $f/v \gg 10$, even this thermal leptogenesis benchmark has twin neutrinos with masses much greater than an MeV. Thus, the twin neutrinos may no longer contribute significantly to $\Delta \Neff$. However, the non-zero mirror photon mass we consider is still useful as it allows the energy density of twin photons to efficiently transfer to the SM before the SM neutrinos decouple, thus preventing a problematic contribution of the twin photon itself to $\Delta \Neff$.

\begin{figure}[t!]
\centering
\includegraphics[width =0.5\textwidth]{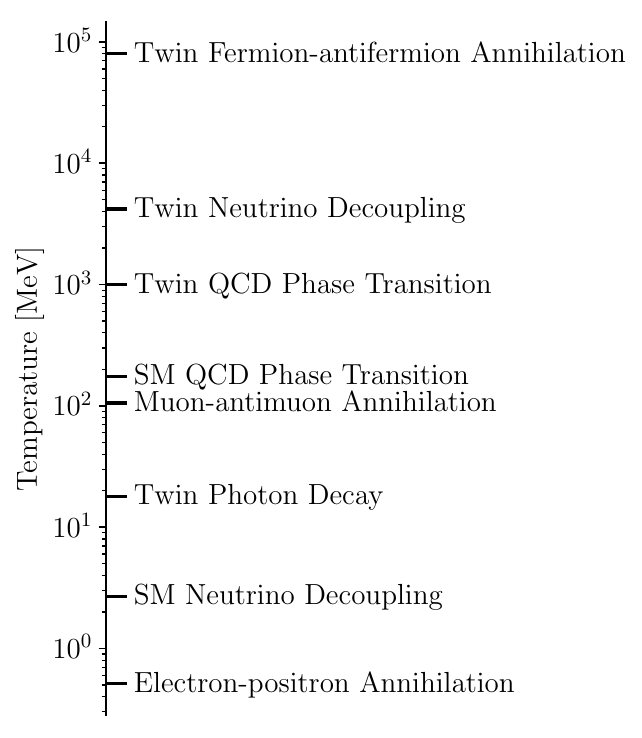}
\caption{\label{fig:timeline} An example cosmic timeline of events in this model that impact observations of $\Neff$.}
\end{figure}

We consider a concrete example where all the charged twin fermion masses are several tens of GeV and where the twin photon has a Stueckelberg mass around $20 \text{ MeV}$. The high mass scale of the twin fermions leads to twin neutrino decoupling before the SM and twin QCD phase transitions. The twin and visible sectors are in thermal contact via kinetic mixing between the twin and SM photons so that entropy can transfer from the twin sector to the SM. The twin and SM QCD phase transitions and SM annihilations heat the SM neutrinos relative to the decoupled twin neutrinos, diluting the twin neutrino contribution to $\Neff$. The cosmic timeline of this scenario is illustrated schematically in Fig.~\ref{fig:timeline}. 
As we will see, with an $18 \text{ MeV}$ twin photon, $\Delta \Neff$ may be as small as $\sim 0.10$. 

The rest of this paper is organized as follows. In Section~\ref{twinsector}, we review the twin Higgs mechanism and determine the necessary MTH mass spectrum for our proposed scenario. We then calculate the twin contributions to $\Delta \Neff$ in Section~\ref{sec:neff} and the effects on the Helium mass fraction in Section~\ref{sec:yp}. Fig.~\ref{fig:NeffvsYp} is the culmination of these calculations which predicts $\mA$, $\Neff$, and the Helium mass fraction for our MTH model. We discuss implications of upcoming experiments and observations and conclude in Section~\ref{sec:conclusion}.

\section{Twin Sector}
\label{twinsector}

\subsection{Twin Higgs mechanism with $\zsym$ breaking}

The MTH model consists of a twin sector that is related to the SM by a $\zsym$ symmetry at a scale above the SM electroweak scale. In particular, the twin sector has a copy of the SM gauge group, $\one' \times \su(2)' \times \su(3)'$, with respective couplings $\prn{g_1',g_2',g_3'}$ and a doublet $H'$ under this $\su(2)'$ which is the twin Higgs. Throughout this paper, superscripts $'$ on SM particles or quantities indicate their twin sector counterparts. An accidental, approximate $\su(4)$ global symmetry in the full Higgs sector is spontaneously broken when the twin Higgs doublet acquires a VEV $f$. The SM Higgs is identified as one of the pseudo-Nambu-Goldstone bosons from the $\su(4)$ breaking whose mass is protected from quadratic divergences by the $\zsym$ symmetry up to the cutoff $\Lambda_{TH}\approx 4\pi f $.  The SM Higgs doublet acquires its measured VEV $v$.

The required fine-tuning (F.T.) of the parameters to obtain the SM electroweak scale $v$ from the twin one $f$ is 
\al{
\label{eq:finetuning}
\text{F.T.}=2\frac{v^2}{f^2}~.
}
The Higgs observed at the LHC has properties that are consistent with the SM Higgs, which places a limit on the ratio of the Higgs VEVs $f \gtrsim 3 v$ \cite{Khachatryan:2016vau}. Thus, tuning in Twin Higgs models is always greater than $\approx 20\%$. In this work, we require that our MTH model does not result in tuning greater than $1\%$ and therefore, that $f/v \lesssim 14$. Requiring our MTH model to be consistent with the latest \emph{Planck} results yields a lower bound of $f/v \gtrsim 10$, as we find below (see Fig.~\ref{fig:NeffvsYp}). In a supersymmetric UV completion of the twin Higgs model with an $\su(4)$ symmetric potential from an $F$ term, fine-tuning of a few percent is already required~\cite{Craig:2013fga}, and $f/v \gtrsim 10$ does not introduce additional fine-tuning. The same is true for a $D$ term model with a high mediation scale of the supersymmetry breaking~\cite{Badziak:2017kjk}.

The Yukawa couplings of the twin and SM sectors may be written succinctly as
\al{
\Lag_{\text{Yuk}} \supset \sum_f - y_{f} \bar{f_R} H f_L - y_{f'} \bar{f'_R} H' f'_L.
} As mentioned, we assume a hard breaking of the $\zsym$ in these Yukawas so that $y_f \ne y_{f'}$ (except for the top Yukawas). In fact, the models we consider have $y_{f'} > y_f$ for all fermions besides the top quarks. We assume that the twin neutrinos are still light and can be treated as dark radiation. We also assume that the gauge coupling constants preserve the $\zsym$ symmetry up to the quantum correction from $\zsym$-breaking fermion masses, which raises the twin QCD scale. We introduce a Stueckelberg mass for the twin photon.

\subsection{Twin Photon} \label{twinphoton}
A crucial requirement for entropy dilution is that the twin photon is able to mediate the transfer of entropy from the twin sector to the SM via the kinetic mixing, 
\al{
\mathcal{L}_{\gamma ' \gamma} = \frac{\epsilon}{2} F'_{\mu \nu} F^{\mu \nu},
}
where $\epsilon$ is the mixing strength between the SM photon and the twin photon, which have field strengths of $F_{\mu \nu}$ and $F'_{\mu \nu}$ respectively. Efficient transfer of entropy is guaranteed as long as the twin photons are thermalized with the SM bath. The twin photons must be massive enough for their decays to proceed in the forward direction at MeV-scale temperatures in order to deplete their number density before BBN. This requirement is satisfied if the twin photon is heavier than a few MeV.

\begin{figure}[t]
\centering
\includegraphics[width =0.5\textwidth]{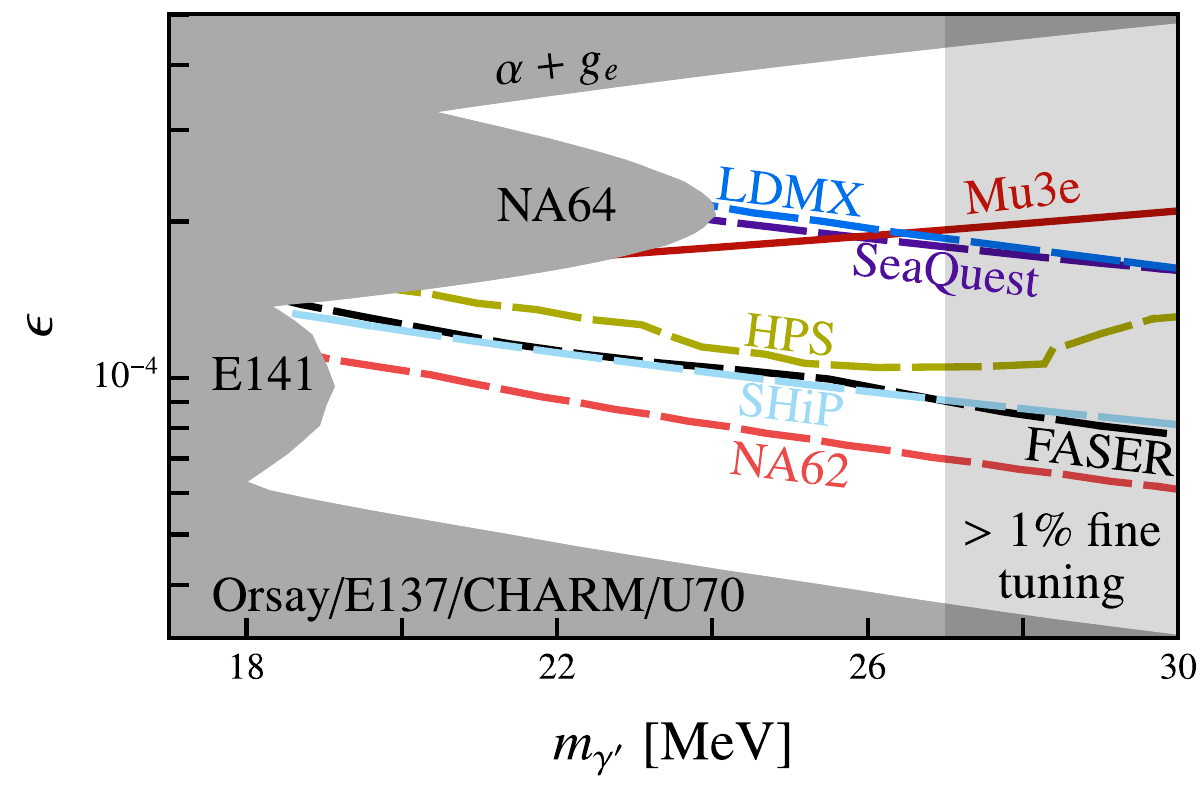}
\caption{\label{fig:mAvseps} Existing and projected constraints on dark photon parameter space. Our MTH model is viable for all values of $\epsilon$ currently unconstrained in the mass range $18 \text{ MeV} \lesssim \mA \lesssim 27 \text{ MeV}$. The beam dump constraints are from the compilation \cite{Alexander:2016aln}, while the $\alpha+g_e$ constraint is from \cite{Parker:2018vye}. The lines are projected constraints from LDMX \cite{Berlin:2018bsc}, Mu3e \cite{Echenard:2014lma}, SeaQuest \cite{Berlin:2018pwi}, HPS \cite{Celentano:2014wya}, SHiP \cite{Alekhin:2015byh}, FASER \cite{Ariga:2018uku}, and NA62 \cite{Lanfranchi:2017wzl}. Dashed lines would rule out the space below the line and solid lines would constrain the space above.}
\end{figure}

In the $1-10$~MeV twin photon mass range, terrestrial and supernova constraints \cite{Chang:2016ntp,DeRocco:2019njg,Sung:2019xie} require $\epsilon \lesssim 10^{-11}$, which is too small to thermalize the twin and SM sectors.  As shown in Fig.~\ref{fig:mAvseps}, larger kinetic mixing is allowed for slightly larger twin photon masses, with constraints from beam dump searches \cite{Alexander:2016aln} and $\alpha+g_e$ measurements \cite{Parker:2018vye} restricting some of the parameter space.
We thus consider twin photon masses above $18$~MeV with values of the kinetic mixing in the range that is allowed by these constraints. The remaining parameter space in $\epsilon$ can be explored with LDMX \cite{Berlin:2018bsc}, Mu3e \cite{Echenard:2014lma}, SeaQuest \cite{Berlin:2018pwi}, HPS \cite{Celentano:2014wya}, SHiP \cite{Alekhin:2015byh}, FASER \cite{Ariga:2018uku}, and NA62 \cite{Lanfranchi:2017wzl}, so this model has considerable discovery potential. Without introducing tuning greater than 1\%, our MTH model is only consistent with cosmological observations for $\mA \lesssim 27 \text{ MeV}$, hence the range of $\mA$ plotted (see Fig.~\ref{fig:NeffvsYp}).

The allowed values of kinetic mixing shown in Fig.~\ref{fig:mAvseps} are more than adequate to thermalize the twin and SM sectors. At high temperatures relative to the twin photon and SM fermion masses, thermalization occurs primarily through $2\to2$ scatters. For example, the rate for $\gamma ' e \to \gamma e$ for temperatures much larger than $\mA$ is roughly
\al{
\Gamma_{\gamma ' e \to \gamma e} \approx \frac{3 \zeta \prn{3}}{8 \pi^3} \epsilon^2 \alpha^2 T,
}
where $\alpha$ is the usual SM fine structure constant. Throughout this paper, $T$ refers to the temperature of the SM photon bath and all SM constants are taken from \cite{Tanabashi:2018oca}. This rate is greater than the Hubble rate for all $T$ in the range $\mA \ll T \lesssim 400 \text{ GeV}$ for the smallest $\epsilon^2 \sim 10^{-9}$ we can consider. For $T \lesssim \mA$, twin photon decays into electron-positron pairs become more efficient. The rest-frame rate for a kinetically mixed twin photon to decay to SM electron-positron pairs is 
\beq \Gamma_{\gamma ' \to e^+ e^-} \approx \frac{ \epsilon^2 \alpha (2 m_e^2 +m_{\gamma'}^2) }{3 m_{\gamma'}}.\eeq 
However, this rate gets suppressed by a factor of $\sim \mA/T$ to account for time dilation at temperatures comparable to or larger than $\mA$. Comparing this rate to the Hubble rate, we find that decays become efficient at mediating entropy transfer below $T\sim 8 \text{ GeV}$ for the smallest $\mA=18 \text{ MeV}$ and $\epsilon^2 \sim 10^{-9}$ we can consider. We conclude that the twin photon can transfer entropy efficiently to the SM for $T\lesssim 400 \text{ GeV}$ for the available parameter space shown in Fig.~\ref{fig:mAvseps}. 

There is an additional, nontrivial requirement on the available twin photon parameter space in Fig.~\ref{fig:mAvseps}: since the twin $Z$ mass eigenstate contains some of the twin photon gauge eigenstate, as discussed in Appendix~\ref{gamp2nubpnup}, the twin $Z$ and SM photon mix. This allows SM fermions to thermally couple to twin neutrinos through elastic scattering and annihilation. For temperatures much larger than the participating SM fermion masses, the cross sections for annihilations $f \bar{f} \to \nu' \bar{\nu}'$ and elastic scatters $\nu ' f \to \nu ' f$ are comparable and roughly
\al{
\sigma_{\nu' f} \approx \frac{16 \pi}{3} \frac{\epsilon^2 Q_f^2 \alpha^2}{\cos^4 \theta_W} \frac{T^2}{m_{Z'}^4}.
}
The total rate for both annihilations and elastic scatters from all SM charged fermions but the top is
\al{
\Gamma_{\nu' f} \approx \frac{640 \zeta(3)}{3\pi} \frac{\epsilon^2 \alpha^2}{\cos^4 \theta_W} \frac{T^5}{m_{Z'}^4}.
}
Ideally, the earliest the twin neutrinos can decouple from the bath is before the SM bottom-antibottom pairs annihilate in our scenario. This rate is smaller than the Hubble rate at $T=m_b$ if
\al{
\epsilon \lesssim 10^{-3}\prn{\frac{f/v}{10}}^2,
}
which is satisfied by the entire parameter space in Fig.~\ref{fig:mAvseps} for our models in which $f/v \gtrsim 10$. Therefore, the effective $Z'-\gamma$~mixing does not re-thermalize the twin neutrinos via scattering with SM fermions.

\subsection{Charged Twin Fermions}

In our setup, the twin neutrinos should decouple from the bath as early as possible. Subsequent QCD phase transitions and SM particle annihilations then raise the temperature of the SM neutrinos relative to the twin neutrinos as much as possible, thus minimizing the twin neutrino contribution to $\Neff$. We consider both the best and next-best scenarios in which the twin neutrinos decouple before the SM bottom-antibottom and SM tau-antitau pairs annihilate, respectively. As we show in Section~\ref{sec:yp}, the best scenario has the largest parameter space consistent with cosmological observations and naturalness (see Fig.~\ref{fig:NeffvsYp}), whereas the next-best scenario requires less $\zsym$-breaking.

For both scenarios, the temperature of twin neutrino decoupling determines the appropriate twin fermion mass spectrum, since elastic scattering off twin fermions is the process that keeps the twin neutrinos in equilibrium at the lowest temperatures. Scattering processes are more important than fermion-antifermion pair annihilations at temperatures below the twin fermion mass since annihilations are suppressed by a relative factor of $e^{-m_{f'}/T}$. The elastic scattering rate is 
\al{
\label{eq:twinnudecpl}
\Gamma_{f' \nu '} \approx \frac{4\prn{3+3\cdot 5}}{\pi} G_F^2 T^2 \frac{v^4}{f^4} \prn{\frac{m_{f'}T}{2\pi}}^{\frac{3}{2}} e^{-m_{f'}/T}.
}
Requiring the elastic scattering rate in Eq.~\eqref{eq:twinnudecpl} to be less than the Hubble rate at $T=m_b$ imposes $m_{f'} \gtrsim 76$ GeV ($m_{f'} \gtrsim 70$ GeV) for $f/v =10$ ($f/v=14$). Thus, for this best scenario where decoupling occurs before $T=m_b$, we set $m_{f'}=80 \text{ GeV}$ for all charged twin fermions besides the twin top. This $\zsym$ symmetry breaking is small enough to not ruin the MTH mechanism.

In the next-best scenario, the twin neutrinos decouple before the SM tau-antitau pairs annihilate. Requiring the elastic scattering rate in Eq.~\eqref{eq:twinnudecpl} to be less than the Hubble rate at $T=m_\tau$ imposes $m_{f'} \gtrsim 27$ GeV ($m_{f'} \gtrsim 24$ GeV) for $f/v =10$ ($f/v=14$). Thus, for this next-best scenario, we set $m_{f'}=30 \text{ GeV}$ for all charged twin fermions besides the twin top. The primary motivation for this next-best scenario is that the $\zsym$ symmetry breaking is even smaller than in the best scenario. Since $m_{f'}=30 \text{ GeV}$, we must additionally consider the Higgs decaying invisibly to twin fermions. The LHC does not probe our predicted Higgs invisible decay rate or reduced Higgs signal strength since we only consider $f/v \gtrsim 10$. However, our predictions for both of these observables fall within the projected capabilities of future colliders such as the ILC \cite{Bambade:2019fyw}, giving another future test of this more $\zsym$-symmetric benchmark. See Appendix~\ref{SMHiggsbnds} for more details. 

\subsection{Twin Gluons} \label{twingluons}
After the charged twin fermions leave the bath, it is still possible for the twin neutrinos to be coupled to the twin gluons. Using the method in Ref.~\cite{Henning:2015alf}, we find that the lowest dimension operator which conserves lepton number and allows twin gluon-neutrino scattering is
\al{
\label{eq:GGnunu} \mathcal{L} \supset
\frac{1}{\prn{4\pi}^4} \frac{1}{m_{q'}^2} \frac{1}{m_{Z'}^2} G^{' a}_{\mu \nu}D_\rho G^{' a \mu \nu} \bar{\nu}' \gamma^\rho \nu',
}
where $G^{' a}_{\mu \nu}$ is the field strength for the twin gluons. Thus, the elastic scattering rate is
\al{
\label{eq:nupgprate}
\Gamma_{\nu ' g' \to \nu ' g'} \approx \frac{1}{\prn{4\pi}^8} \frac{1}{m_{q'}^4} \frac{1}{m_{Z'}^4} T^9.
}
For the best scenario, requiring that the rate in Eq.~\eqref{eq:nupgprate} is less than the Hubble rate at $T=m_b$ yields
\al{
m_{q'} &\gtrsim 2 \prn{\frac{f/v}{10}} \text{ GeV},
}
which is easily satisfied because $m_{f'}=80 \text{ GeV}$. For the next-best scenario, requiring the rate in Eq.~\eqref{eq:nupgprate} is less than the Hubble rate at $T=m_\tau$ yields an even more trivially satisfied condition for our benchmark $m_{f'}=30 \text{ GeV}$. Thus, $\nu'-g'$ scattering does not re-thermalize the twin neutrinos.

In order for the entropy in the twin gluons to be transferred to the SM via twin photons, we require that the twin photons and gluons stay in equilibrium after the twin quarks leave the thermal bath and as the twin QCD phase transition is proceeding. Integrating out the heavy twin quarks, the twin gluons and photons are coupled at lowest order by the dimension-8 operators \cite{Novikov:1977dq}
\al{
\Lag_{F'F'G'G'} &= \frac{\alpha' \alpha_S'}{180} \prn{ \frac{\sum_i Q_i^2}{m_{q'}^4} } \left[ 28 F'_{\mu \nu}F'_{\nu \lambda} G^{' a}_{\lambda \sigma}G^{' a}_{\sigma \mu}+14F'_{\mu \nu}F'_{\lambda \sigma} G^{' a}_{\sigma \mu}G^{' a}_{\nu \lambda} \right. \nonumber \\
&\left. -10 \prn{F'_{\mu \nu} G^{' a}_{\mu \nu}} \prn{F'_{\alpha \beta} G^{' a}_{\alpha \beta}}-5 \prn{F'_{\mu \nu} F'_{\mu \nu}} \prn{G^{' a}_{\alpha \beta}G^{' a}_{\alpha \beta}} \right] 
}
where $\alpha',\alpha_S'$ are the twin U$(1)_{\text{EM}}'$ and SU$(3)'$ fine structure constants. We sum over the twin quark charges-squared (aside from the top, which contributes negligibly). We require that the $2\to2$ scattering rate provided by this coupling is faster than the Hubble rate at the twin QCD phase transition
\al{
\label{eq:gluoncond}
\evat{H}_{\Lambda_\text{QCD}'} &\lesssim 0.01 \prn{\frac{\alpha' \alpha_S'}{m_{q'}^4}}^2 \Lambda_{\text{QCD}}'^9 
&\Longrightarrow& &m_{q'} \lesssim 100 \prn{\frac{\Lambda_\text{QCD}'}{2 \text{ GeV}}}^{7/8} \text{ GeV}.
}
Here we take $\alpha_S' \prn{\Lambda_\text{QCD}'}=4\pi$, but the upper bound on $m_{q'}$ weakly depends on the value.
For $f/v=10$ and $m_{q'}= 30-80$ GeV, we find $\Lambda_\text{QCD}'=1.8-2.5$ GeV. 
This condition is satisfied by $m_{f'}=80 \text{ GeV}$ ($m_{f'}=30 \text{ GeV}$) in the best (next-best) scenario. Hence, the twin photons and twin gluons are in equilibrium throughout the twin QCD phase transition and the entropy is transferred to the twin photons and therefore the SM bath efficiently. We obtain the same conclusion by computing the decay rate of twin glueballs into a pair of twin photons.

\section{Twin Contributions to \texorpdfstring{$\Delta \Neff$}{Delta Neff}}
\label{sec:neff}
We have established that the twin neutrinos decouple before the SM bottom-antibottom pairs leave the bath in the best-case scenario. The particles in the twin and SM thermal bath after twin neutrino decoupling are:
\begin{itemize}
\item SM and twin gluons and photons
\item all SM quarks, except the top
\item all SM leptons. 
\end{itemize}
In the next-best scenario, the twin neutrinos decouple before the SM tau-antitau pairs annihilate and the particles in the bath after twin neutrino decoupling are the same except for the absent SM bottoms.
We also established that thermalization of the SM and twin baths is guaranteed by twin photons for all temperatures $T \lesssim 400 \text{ GeV}$. 

Entropy in the bath is given by
\al{
\label{eq:sdefn}
s= \frac{2\pi^2}{45} g_{\ast s} T^3,
}
where $g_{\ast s}(T)$ tracks the effective number of relativistic degrees of freedom. After particles in the twin and SM sectors annihilate or decay, their entropy cascades down to lighter species that are still coupled. Conservation of entropy then requires that the relative temperature between the twin neutrinos and the thermal bath is 
\beq \left(\frac{T_{\nu'}}{T_b}\right)^3 = \frac{g_{*s}}{g_{*s, 0}},\eeq 
where $g_{*s,0}$ is the effective number of degrees of freedom still in the thermal bath just after twin neutrino decoupling and $g_{*s}$ is the effective number of degrees of freedom at some later time. 
At the time of SM neutrino decoupling, $T_{\nu}^{\dec} \approx 2.7$ MeV \cite{Mangano:2006ar}, $g_{*s} = 43/4$.  Meanwhile, given the degrees of freedom listed above, $g_{*s, 0} = 421/4$ in the best scenario and $g_{*s, 0} = 379/4$ in the next-best. The smallest possible contribution to $\Delta \Neff$ from twin neutrinos occurs when they do not receive any entropy injections after decoupling from the twin bath. Assuming this happens and using the definition in Eq.~\eqref{eq:Neffdefn}, the contribution of twin neutrinos to $\Neff$ is
\beq 
\label{eq:DNeffmin}
\Delta \Neff^{\nu',\text{min.}} = 3 \left(\frac{43/4}{421/4}\right)^{4/3} \approx 0.14
\eeq
for the best scenario and $0.16$ for the next-best. 
Thus, both scenarios seem allowed by the latest \emph{Planck} results~\cite{planck2018} which give $\Delta \Neff=\Neff-3.046<0.284$ at 95\% confidence.

However, twin photons decay into twin neutrinos since the twin photon mass eigenstate has a small amount of the twin $Z$ gauge eigenstate (see Appendix~\ref{gamp2nubpnup}). Thus, the $\Delta \Neff^{\nu',\text{min.}}$ in Eq.~\eqref{eq:DNeffmin} is never attainable in practice. To account for this reheating of the decoupled twin neutrinos, we must solve their energy density Boltzmann equation
\al{
\label{eq:BE1}
\dd_t \rho_{\nu '} +4H \rho_{\nu '}=m_{\gamma '} \Gamma_{\gamma ' \rightarrow \bar{\nu}' \nu '} n_{\gamma '}^{\text{eq}} \prn{T},
}
where $\rho_{\nu '}$ is the total energy in all 3 twin neutrino species and $\Gamma_{\gamma ' \rightarrow \bar{\nu}' \nu '}$ is the total decay rate of $\gamma '$ into any of the twin neutrino pairs, given by Eq.~\eqref{eq:Ggamp2nup} in Appendix~\ref{gamp2nubpnup}. The twin photons are in chemical and kinetic equilibrium with the SM with the number density \beq n_{\gamma '}^{\text{eq}}(T)= \frac{3 \mA^2 T}{2\pi^2} K_2(\mA/T).\eeq We ignore the the back reaction and neutrino Pauli blocking in Eq.~\eqref{eq:BE1} since the number density of twin neutrinos is small in order for $\Delta \Neff$ to be consistent with observations. By neglecting inverse decays, we overestimate $\rho_{\nu'}$ and therefore
overestimate the twin contribution to $\Delta \Neff$. With the change of variables $\rho_{\nu '} \equiv s^{4/3} y$, Eq.~\eqref{eq:BE1} simplifies to 
\al{
\label{eq:dydT}
\frac{\dd y}{\dd T}=-\frac{m_{\gamma '} \Gamma_{\gamma ' \rightarrow \bar{\nu}' \nu '} n_{\gamma '}^{\text{eq}} \prn{T}}{3Hs^{4/3}}\prn{\frac{3}{T}+ \frac{\dd g_{\ast s}}{\dd T} \frac{1}{g_{\ast s}}}.
}
For the range of $m_{\gamma'}$ we consider around tens of MeV, we find that $\frac{\dd g_{\ast s}}{\dd T} \frac{1}{g_{\ast s}} \ll \frac{3}{T}$.
We integrate \eqref{eq:dydT} to find
\al{
\label{eq:ysoln}
y(T)-y(T_0)=\frac{c_y \Gamma_{\gamma ' \rightarrow \bar{\nu}' \nu '} M_\text{Pl}}{\mA^2} \int_{\mA/T_0}^{\mA/T}dx \frac{x^4 K_2(x)}{\sqrt{g_\ast} g_{\ast s}^{4/3}}=\frac{c_y c_\text{int} \Gamma_{\gamma ' \rightarrow \bar{\nu}' \nu '} M_\text{Pl}}{\mA^2},
}
where \beq x\equiv \frac{\mA}{T}, \quad c_y \equiv \frac{9 \sqrt{10}}{2 \pi^3 \prn{\frac{2\pi^2}{45}}^{4/3}},\quad
\text{and  } c_\text{int}=0.26 \eeq is the value of the dimensionless integral.
We evaluate the integral from $x_i=1/5$ to $x_f=10$ because it effectively converges over this domain and the twin photons have all but left the bath by $x_f$. The integral doesn't change appreciably over our range of $\mA$.

We thus find the final energy density, $\rho_{\nu'}$,
\al{
\label{eq:rhonupf}
\rho_{\nu '} \prn{T_f} \!=\! \evat{s^{4/3}y}_{T_f} =T_f^4 \prn{\! \frac{\pi^2}{30}\frac{7}{8}6 \prn{\frac{\evat{g_{\ast s}}_{T_f}}{\evat{g_{\ast s}}_{T_{\nu '}^{\dec}}}}^{4/3} \right.  \!+\! \left. \prn{\frac{2\pi^2}{45} \evat{g_{\ast s}}_{T_f}}^{4/3}\frac{c_y \Gamma_{\gamma ' \rightarrow \bar{\nu}' \nu '} M_\text{Pl}}{\mA^2} c_{\text{int}}\!} \!.
}
We translate this energy density into the corresponding contribution to $\Delta \Neff$.
At $T_f$, the energy density in a single SM neutrino is just $\frac{7}{4} \frac{\pi^2}{30} T_f^4$. Even though the SM neutrinos may have decoupled before $T_f=\frac{m_{\gamma '}}{10}$, they are still at the same temperature as the SM bath since electron-positron pairs do not start to annihilate in the forward direction until $T\lesssim 1$ MeV and the smallest $T_f$ we consider is $T_f=\frac{18 \text{ MeV}}{10}=1.8 \text{ MeV}$. Taking the ratio of the final twin-neutrino energy density from~\eqref{eq:rhonupf} to a single SM neutrino's, we find
\al{
\label{eq:DNeffNup}
\Delta \Neff^{\nu '}=3 \prn{\frac{\evat{g_{\ast s}}_{T_f}}{\evat{g_{\ast s}}_{T_{\nu '}^{\dec}}}}^{4/3} + \prn{\evat{g_{\ast s}}_{T_f}}^{4/3} \frac{540 \sqrt{10} c_{\text{int}}}{7 \pi^5} \frac{\Gamma_{\gamma ' \rightarrow \bar{\nu}' \nu '} M_\text{Pl}}{\mA^2} .
}
This simplifies to the result in Eq.~\eqref{eq:DNeffmin}, in the limit $\Gamma_{\gamma ' \rightarrow \bar{\nu}' \nu '} \rightarrow 0$.

There is still an appreciable number density of twin photons in the SM thermal bath when the SM neutrinos decouple. These twin photons subsequently decay to electron-positron pairs with which they are in equilibrium. This causes the SM neutrinos to be cooler than usual relative to the SM photons. Thus, the twin photons contribute negatively to $\Delta \Neff$, denoted by $\Delta \Neff^{\gamma'}$. Using entropy conservation at $T_\nu^{\dec}$ and $T_f$, we find: 
\al{
\label{eq:TNuovrT}
\frac{T_\nu}{T}&=\prn{\frac{g_{\ast s}\prn{T_f}}{g_{\ast s}\prn{T_\nu^{\dec}}}}^{1/3}
=\prn{\frac{4}{11+2g_{\ast s}^{\gamma '} \prn{T_\nu^{\dec}}}}^{1/3}.
}
Comparing the energy density at this reduced temperature to the definition of $\Neff$ in Eq.~\eqref{eq:Neffdefn}, we find
\al{
\label{eq:DNeffNu}
\Delta \Neff^{\gamma'} = 3\cdot \prn{\frac{11}{11+2g_{\ast s}^{\gamma '} \prn{T_\nu^{\dec}}}}^{4/3} -3.
}
Of course, SM neutrinos do not decouple instantaneously at $2.7 \text{ MeV}$. Some of the entropy transfer from these dark photon decays into SM electron-positron pairs will eventually move into SM neutrinos so that their temperature relative to the SM photons is not quite as small as in \eqref{eq:TNuovrT}. This should not introduce more than a $10\%$ error in our $\Delta \Neff^{\gamma'}$ calculation.
Combining the $\Delta \Neff$ contribution in~\eqref{eq:DNeffNu} with the contribution in~\eqref{eq:DNeffNup}, we arrive at our final change to $\Neff$
\al{
\Delta \Neff = \Delta \Neff^{\nu '}+ \Delta \Neff^{\gamma'}.
}

\section{The Helium Mass Fraction}
\label{sec:yp}
For twin photon masses as light as 18 MeV to be consistent with measurements of $\Neff$, the negative contribution to $\Delta \Neff$ from $\gamma'$ decay in Eq.~\eqref{eq:DNeffNu} is critical. This change in the ratio between SM photon and neutrino temperatures occurs close to the time of BBN and thus may affect the primordial Helium mass fraction $Y_P$, which has been measured to be $Y_P=0.2449 \pm 0.0040$ \cite{Aver:2015iza}. This observable is sensitive not only to the expansion rate at BBN but also to the weak interaction rates, which are themselves dependent on the electron neutrino temperature relative to the photon bath. Since the decaying twin photons alter this ratio of temperatures, we must ensure our prediction for $Y_P$ is consistent with measurement.

Our analysis relies on the numerical results from Ref.~\cite{Galvez:2016sza} which uses a modification of the publicly available \texttt{AlterBBN} code~\cite{Arbey:2011nf,Arbey:2018zfh}. Ref.~\cite{Galvez:2016sza} calculates cosmological observables as a function of the number of degrees of freedom that are relativistic at recombination (besides SM photons) as well as the effective temperature of those degrees of freedom. They refer to these degrees of freedom as neutrinos since they include SM neutrinos. But, since they vary both the number, $N_\nu$, and temperature, $T_\nu$, of these degrees of freedom, their parameterization subsumes our situation in which we change the temperature of SM neutrinos and have extra relativistic degrees of freedom at BBN. We calculate $T_\nu$ relative to its usual temperature in the SM using Eq.~\eqref{eq:TNuovrT}
\al{
\label{eq:TNu}
\frac{T_\nu}{T_{\nu \SM}}=\prn{\frac{1}{1+\frac{2}{11}g_{\ast s}^{\gamma '} \prn{T_\nu^{\dec}}}}^{1/3}.
}
$N_\nu$ as defined in Ref.~\cite{Galvez:2016sza} is related to $\Neff$ by
\al{
\Neff T_{\nu \SM}^4 = N_\nu T_\nu^4,
}
since the $\Neff$ which appears in Eq.~\eqref{eq:Neffdefn} is inferred from measurements of the total energy density at BBN and recombination.
From this relation, we find
\begin{figure}[t]
\includegraphics[width =0.49\textwidth]{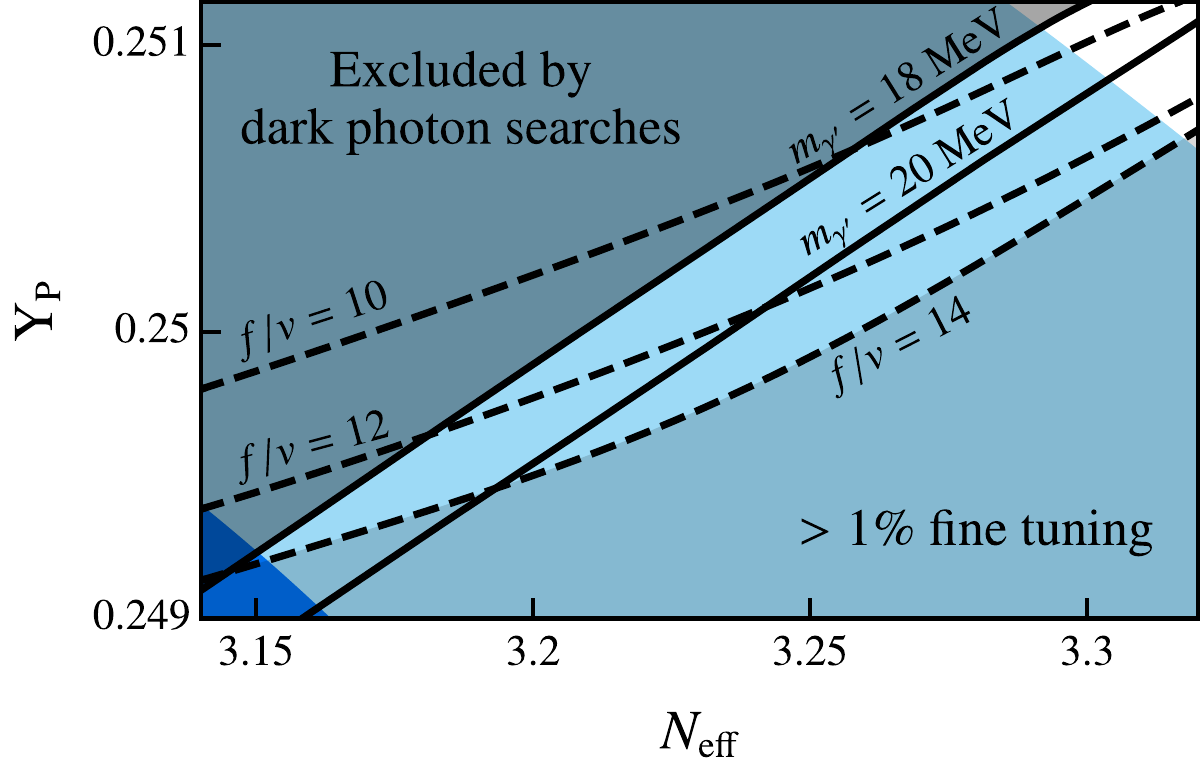}
\includegraphics[width =0.49\textwidth]{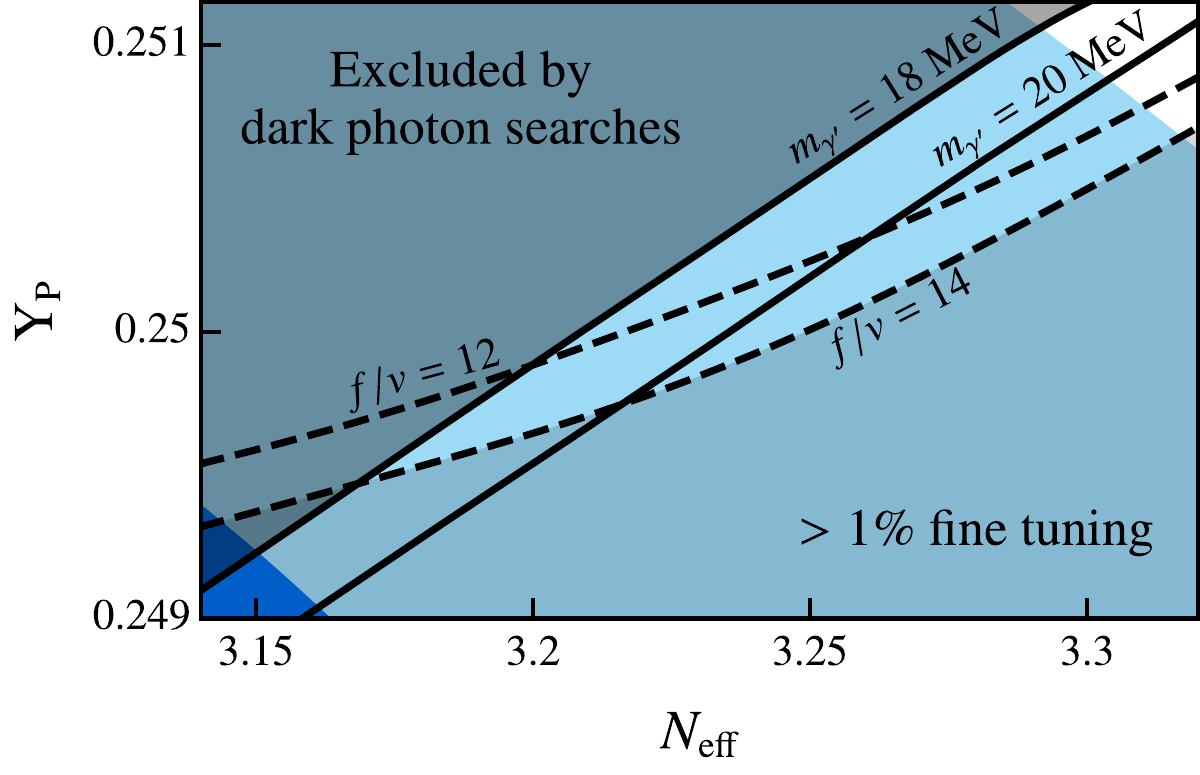}
\caption{\label{fig:NeffvsYp} Contours of constant $\mA$ (solid) and $f/v$ (dashed) on the $\Neff$-$Y_P$ plane assuming $m_{f'}=80$ (left) and 30 (right) GeV for non-top, charged twin fermions. The dark and light blue regions are respectively the $1 \sigma$ and $2\sigma$ containment from \emph{Planck}~\cite{planck2018}. They combine the \emph{Planck} TT, TE, and EE+lowE+lensing+BAO data with the $Y_P$ bounds from \cite{Aver:2015iza}. Twin photons lighter than $18$~MeV are constrained by experiments and $f/v\gtrsim 14$ requires fine-tuning greater than 1\%.}
\end{figure}
\al{
\label{eq:NNu}
N_\nu = 3.046+\Delta \Neff^{\nu '} \prn{\frac{T_{\nu \SM}}{T_\nu}}^4,
}
where the first term is the contribution from SM neutrinos~\cite{deSalas:2016ztq,Mangano:2001iu} and the second term from twin neutrinos, as in Eq.~\eqref{eq:DNeffNup}. With Eqs.~\eqref{eq:TNu} and~\eqref{eq:NNu}, we use the results of Ref.~\cite{Galvez:2016sza} to calculate $Y_P$.

The left panel of Fig.~\ref{fig:NeffvsYp} shows contours of $\mA$ and $f/v$ on the $\Neff$-$Y_P$ plane for the best scenario in which we set $m_{f'}=80 \text{ GeV}$ so that the twin neutrinos decouple before the SM bottom-antibottom pairs annihilate.
Additionally, we include the $1 \sigma$ and $2\sigma$ containment from \emph{Planck}~\cite{planck2018} as dark and light blue regions, respectively, resulting in a slim parameter space where both the cosmology and naturalness of these models is reasonable.
For the lightest twin photon we can consider, $\mA=18$~MeV, the data require that $f/v \gtrsim 10$.
For larger $\mA$, larger $f/v$ are necessary to suppress the twin photon decays to twin neutrinos. In order to have a twin photon as heavy as $\mA=27$~MeV, $f/v \gtrsim 14$ is required. The smallest $\Delta \Neff$ we can achieve is $0.10$ and corresponds to $\mA=18 \text{ MeV}$ and $f/v=14$.

The right panel of Fig.~\ref{fig:NeffvsYp} is equivalent for the next-best scenario in which we set $m_{f'}=30 \text{ GeV}$ so that the twin neutrinos decouple before the SM tau-antitau pairs annihilate. Again, the lightest twin photon $\mA=18$~MeV requires $f/v \gtrsim 10$, but the largest $\mA$ consistent with cosmological data when $f/v=14$ is $25 \text{ MeV}$. The smallest $\Delta \Neff$ we can achieve for this scenario is $0.12$ and again corresponds to $\mA=18 \text{ MeV}$ and $f/v=14$. 

\section{Discussion}
\label{sec:conclusion}
In this paper, we have considered a new way to mitigate the $\Neff$ problem of MTH models. While other works have considered lifting twin Yukawa couplings as we have done here, we have additionally given the twin photon a mass. This greatly reduces $\Delta \Neff$ by allowing all of the entropy transferred after the twin neutrinos decouple to eventually go into the SM bath instead of staying in the twin photons. In the best scenario, all charged twin fermions (besides the twin top) have $m_{f'}=80 \text{ GeV}$. For this spectrum, the twin neutrinos decouple before the SM bottom-antibottom pairs leave the bath which yields our smallest possible $\Delta \Neff=0.10$ when $\mA=18 \text{ MeV}$ and $f/v=14$. We have carefully accounted for the effects of the twin spectrum not only on $\Neff$ but also on $Y_P$ when determining the viability of our model. We also considered the next-best scenario in which $m_{f'}=30 \text{ GeV}$ so that the twin neutrinos decouple before the SM tau-antitau pairs leave the bath. For this scenario, the smallest possible $\Delta \Neff=0.12$ corresponds to $\mA=18 \text{ MeV}$ and $f/v=14$. One simple generalization of both mass benchmarks would be to allow a smaller twin hypercharge gauge coupling. This would decrease the rate of twin photon decays into twin neutrinos and therefore $\Delta \Neff$. However, as our motivation has been to maintain minimal $\zsym$ breaking, we do not pursue this further here.

CMB stage~3 experiments \cite{Benson:2014qhw,Louis:2016ahn,Suzuki:2015zzg,Grayson:2016smb} are projected to reach a sensitivity of $\Delta\Neff \sim 0.06$, while stage~4 experiments have a target $\Delta\Neff = 0.027$ \cite{CMBS4}. Fig.~\ref{fig:NeffvsYp} shows that $\Delta \Neff \gtrsim 0.10$ and $\Delta \Neff \gtrsim 0.12$ in our heavier and lighter MTH models, making them imminently discoverable by current and future observations. Current experimental constraints and naturalness considerations allow $\mA \in \left[18,27 \right] \text{ MeV}$ with a kinetic mixing $\epsilon \sim \mathcal{O} \prn{10^{-4}}$. Interestingly, this parameter space is also imminently discoverable by a host of proposed experiments, as shown in Fig.~\ref{fig:mAvseps}. Whether from CMB light or dark-photon light, we will soon know if our MTH model is viable and accurately predicts an observable $\Delta\Neff$ and massive dark photon.

\acknowledgments
We thank Seth Koren and Takemichi Okui for helpful discussions. We acknowledge the importance of equity and inclusion in this work and are committed to advancing such principles in our scientific communities. KH is supported by the U.S. DOE under Contract DE-SC0009988. HM is supported by the U.S. DOE under Contract DE-AC02-05CH11231, and by the NSF under grant PHY-1638509. HM is also supported by the JSPS Grant-in-Aid for Scientific Research (C)
(No.~17K05409), MEXT Grant-in-Aid for Scientific Research on Innovative Areas (No. 15H05887, 15K21733), by WPI, MEXT, Japan, and Hamamatsu Photonics K.K. RM and KS are supported by the National Science Foundation Graduate Research Fellowship Program. KS is also supported by a Hertz Foundation Fellowship.

\appendix
\section{\texorpdfstring{$\gamma ' \to \bar{\nu}' \nu '$}{Dark Photon to Dark Neutrino} Decays} \label{gamp2nubpnup}

The rate of $\gamma ' \to \bar{\nu}' \nu '$ depends on the amount of $\gamma '-Z'$ mixing. The relevant parts of the twin Lagrangian are
\al{
\Lag^{\text{twin}} \supset -\frac{1}{4} \prn{W_{\mu \nu}'^3}^2 -\frac{1}{4} \prn{B_{\mu \nu}'}^2
+\frac{1}{2} m_D^2 B_\mu'^2+\frac{1}{2}m_{Z'}^2 Z_\mu'^2,
}
where $m_D$ is the mass of the twin hyper charge gauge boson. Using the weak-angle rotation, we find these terms may be written as
\al{
\label{eq:twin_gbasis}
\Lag^{\text{twin}} \supset -\frac{1}{4} \prn{Z_{\mu \nu}'}^2 -\frac{1}{4} \prn{F_{\mu \nu}'}^2 +\frac{1}{2}\pmat{Z'_\mu& A'_\mu}\pmat{m_{Z'}^2+s_{W'}^2 m_D^2& -s_{W'}c_{W'}m_D^2 \\ -s_{W'}c_{W'}m_D^2& c_{W'}^2 m_D^2}\pmat{Z'_\mu \\ A'_\mu}, 
}
where $c_{W'} \equiv \cos \theta_{W'}$ and $s_{W'} \equiv \sin \theta_{W'}$.
When $m_D^2=0$, $\prn{Z',A'}$ is just the normal twin mass basis. The eigenvalues of the symmetric mass-squared matrix in~\eqref{eq:twin_gbasis} are
\al{
\label{eq:ApZpmsq}
&\mA^2=m_D^2c_{W'}^2-\ord{\frac{m_D^2}{m_{Z'}^2}}, & & &m_{\tilde{Z}'}^2=m_{Z'}^2+\ord{\frac{m_D^2}{m_{Z'}^2}}.
}
The mass matrix is rotated to the mass basis $\prn{\tilde{Z}'_\mu,\tilde{A}'_\mu}$ via
\al{
\label{eq:Zpgampmassrot}
\pmat{Z'_\mu\\ A'_\mu}=\pmat{\cos\theta& +\sin\theta \\ -\sin\theta& \cos\theta}\pmat{\tilde{Z}'_\mu \\ \tilde{A}'_\mu},
}
where $ \cos\theta=1-\ord{m_D^4/m_Z'^4}$ and \beq \sin\theta=s_{W'} c_{W'} \frac{m_D^2}{m_{Z'}^2}+\ord{\frac{m_D^4}{m_{Z'}^4}}
=\frac{s_{W'}}{c_{W'}} \frac{\mA^2}{m_{Z'}^2}+\ord{\frac{m_D^4}{m_{Z'}^4}}.\eeq
The mass eigenstate twin photon $\gamma '$ has a small mixing with the gauge eigenstate $Z'$ given by
\al{
\label{eq:ZpApmix}
\sin\theta=\frac{s_{W'}}{c_{W'}} \frac{m_{\gamma '}^2}{m_{Z'}^2}.
}
The decay rate of the $Z$ boson to a single generation of neutrinos in the SM
\al{
\Gamma_{Z \to \bar{\nu} \nu}=\frac{\alpha M_Z}{24 s^2_W c^2_W},
}
where we neglected the $\nu$ masses.
Since the twin photon mixes with the twin Z, the total decay rate is
\al{
\label{eq:Ggamp2nup}
\Gamma_{\gamma ' \to \bar{\nu}' \nu '}=\frac{\alpha' m_{\gamma '}}{8 s^2_{W'} c^2_{W'}} \sin^2\theta =\frac{\alpha' }{8 c^4_{W'}} \frac{m_{\gamma '}^5}{m_{Z'}^4} =\frac{g'^{ 2}_1}{2\pi g'^{ 2}_2 \prn{g'^{ 2}_2+g'^{ 2}_1}}\frac{m_{\gamma '}^5}{f^4}. 
}
To minimize $\zsym$-breaking, we take $\alpha ' = \alpha$, $\cos \theta_{W'}=\cos \theta_{W}$, and $m_{Z'}=f/v \cdot m_{Z}$.

\section{Higgs Invisible Decays and Signal Strength} \label{SMHiggsbnds}
\begin{figure}[t]
\includegraphics[width =0.49\textwidth]{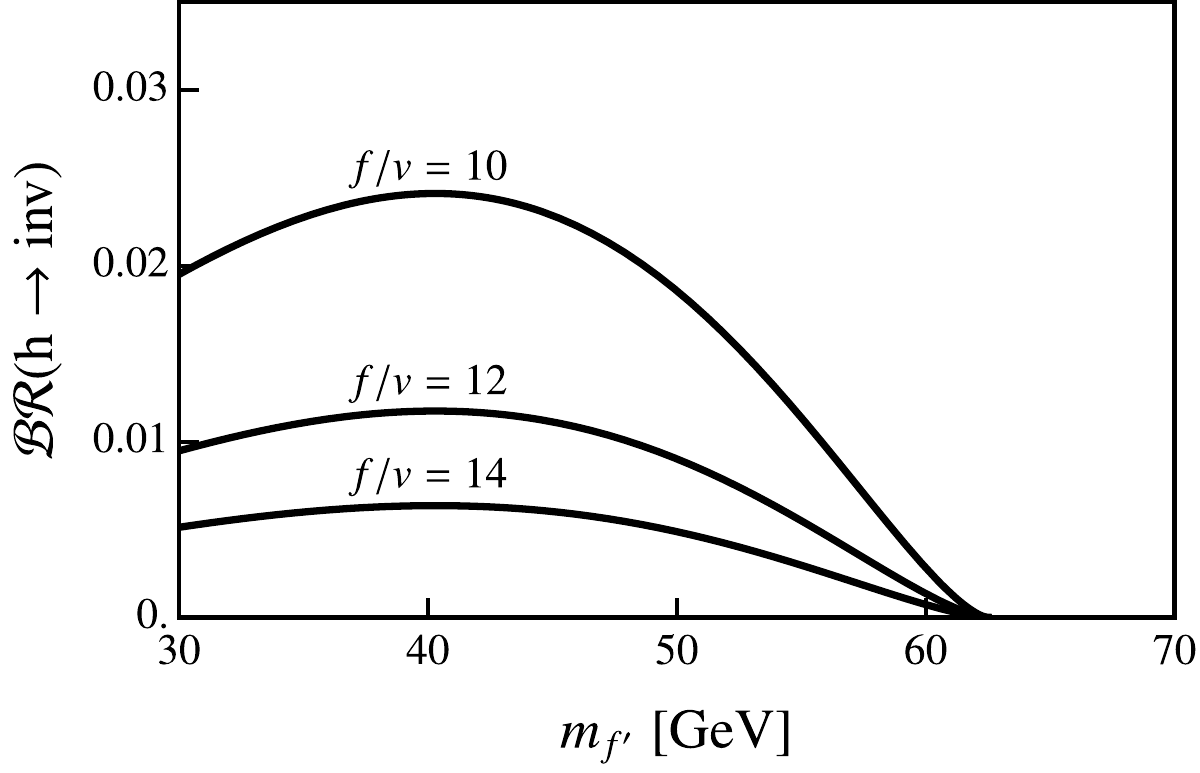}
\includegraphics[width =0.49\textwidth]{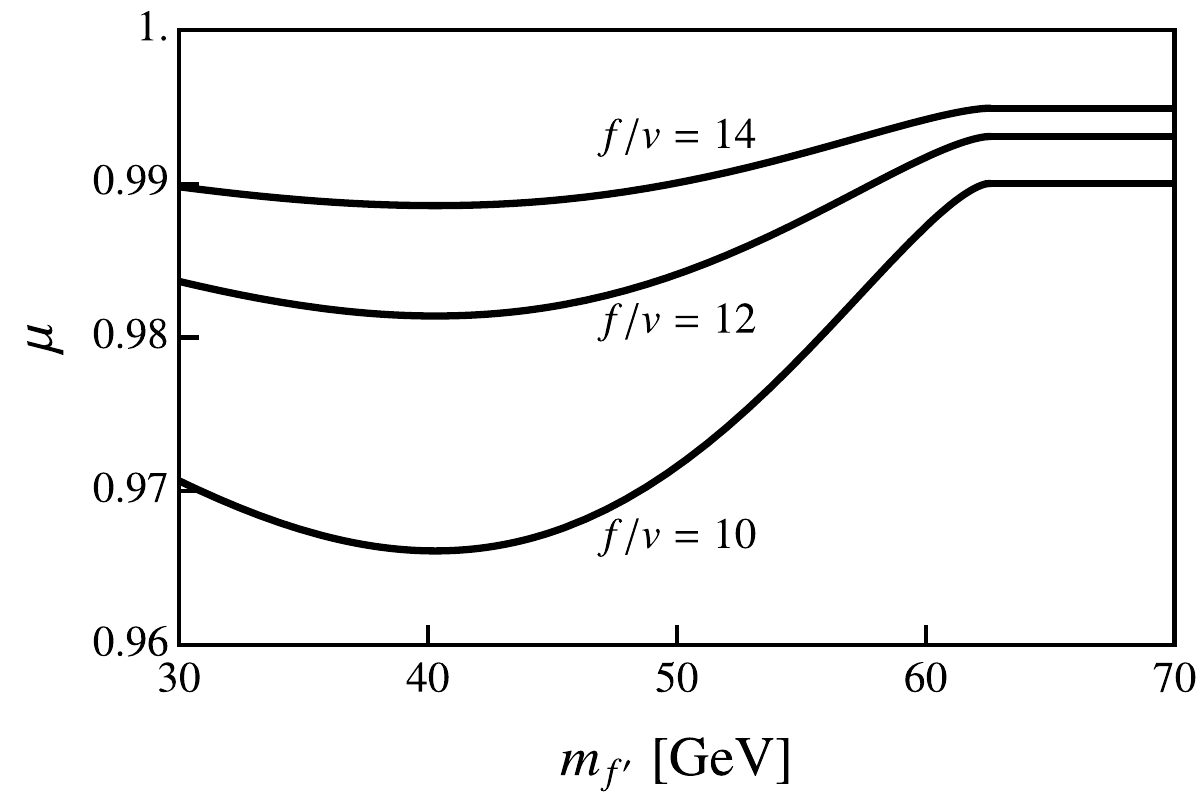}
\caption{\label{fig:BRhtoinv} Higgs-to-invisible branching ratio (left) and Higgs signal strength (right) as a function of $m_{f'}$ for various $f/v$. 
} 
\end{figure}

In Twin Higgs models, the SM-like Higgs we observe decays to invisible twin particles because the SM-like Higgs, $h$, is a mixture of both the physical SM Higgs, $h_{\rm phys}$, and the physical twin Higgs, $h'_{\rm phys}$: 
\al{
\label{eq:hmix}
h=\cos \left(v / f\right)h_{\rm phys}+\sin \left(v / f\right)h'_{\rm phys} \approx \prn{1-1/2 \prn{v/f}^2}h_{\rm phys}+v/f\cdot h'_{\rm phys},
}
where the approximation in the second line is valid for the $f/v \gtrsim 10$ we consider.
Twin fermions couple to the twin Higgs
with coupling $\frac{y_{f'}}{\sqrt{2}}=\frac{m_{f'}}{f}$.
The total SM-like Higgs decay rate to ``invisible'' twin fermions is
\al{
\label{eq:gam h to inv}
\Gamma^{\inv}_h \!=\! \frac{m_{h} v^4}{8\pi f^4} \prn{1\!-\!\left(\frac{2m_{f'}}{m_{h}}\right)^2}^{3/2}\Bigg[ N_{l'}\frac{m_{f'}^2}{v^2}  \!+\! 3N_{q'}\left(\frac{m_{q'}\left(m_{h}\right)}{v}\right)^{2} \left(1\!+\!\frac{5.67}{\pi} \alpha_{S'}(m_{h}) \right) \Bigg],
}
where $N_{q'}$ is the number of twin quarks that the Higgs can decay into and $N_{l'}$ is the number of twin leptons it can decay into. While the tree-level rate is sufficiently accurate for twin leptons, we must include twin QCD radiative and running-quark-mass corrections in the decay rate into twin quarks. We set $\alpha_{S'}\prn{m_h}=\alpha_{S}\prn{m_h}=0.112$ \cite{Tanabashi:2018oca}, as is roughly required by the TH mechanism. The running quark mass to leading order is \cite{Tanabashi:2018oca}
\al{
m_{q'} \prn{m_h}=m_{q'}\prn{1-\frac{\alpha_{S'}\prn{m_h}}{\pi}\prn{\frac{4}{3}+\log \frac{m_h^2}{m_{q'}^2}}}.
}

The total decay width of the SM Higgs with $m_h=125$ GeV is $\Gamma^{\SM}_h=4.07\times 10^{-3}$ GeV, with a relative uncertainty of $\approx 4 \%$ both up and down~\cite{Denner:2011mq}. 
Thus, we only require our own theoretical uncertainties in $\Gamma_h^{\inv}$ to be less than $\approx 10 \%$. 
Note that the total Higgs decay rate, $\Gamma_h$, is related to the total Higgs decay rate in the SM, $\Gamma^{\SM}_h$, via $\Gamma_h=\prn{1-\prn{v/f}^2} \Gamma^{\SM}_h+\Gamma_h^{\inv}$. 
Thus, we find the Higgs-to-invisible branching ratio
\al{
\mathcal{BR}\left(h\rightarrow \inv\right)=
\frac{\Gamma_h^{\inv}}{\Gamma_h}=
\frac{\Gamma_h^{\inv}}{\prn{1-\prn{v/f}^2} \Gamma^{\SM}_h+\Gamma_h^{\inv}}
=\left(1+\frac{\prn{1-\prn{v/f}^2} \Gamma^{\SM}_h}{\Gamma_h^{\inv}}\right)^{-1},
}
where $\Gamma_h^{\inv}$ is given by Eq.~\eqref{eq:gam h to inv}. 
We require the branching ratio to anything in the twin sector to total less than $0.25$ \cite{Tanabashi:2018oca}. 

Fig.~\ref{fig:BRhtoinv} demonstrates that our light twin benchmark with $m_{f'}=30 \text{ GeV}$ and $f/v \gtrsim 10$ is well below the current invisible branching ratio bound.
The $250 \text{ GeV} $ ILC  will be able to probe Higgs invisible decays down to $0.3\%$ \cite{Bambade:2019fyw}. Incredibly, the ILC will therefore be able to probe the entire $f/v$ parameter space for our light twin benchmark.

In addition to evading the current limit on the Higgs-to-invisible branching ratio, 
we also need our light twin benchmark to satisfy bounds on the Higgs signal strength. 
We define the Higgs signal strength as \cite{ATLAS:2017ovn}
\begin{align}
\label{eq:mu defn}
\mu = \frac{\sigma \times \mathcal{BR}}{\left(\sigma \times \mathcal{BR}\right)_{\SM}}.
\end{align}
Since the SM-like Higgs is not quite the SM Higgs, any cross section which yields a single Higgs in the final states will be suppressed by the same amount, namely 
\begin{equation}
\frac{\sigma}{\sigma_{\SM}}=1-\prn{v/f}^2
\label{eq:sigmas ratio}
\end{equation}
Additionally, the Higgs branching ratio for any Higgs decay to SM particles, $h\rightarrow f$, will be reduced. 
The Higgs decay rate itself will be reduced by the same factor as the production cross section
$\Gamma_{h\rightarrow f}=\prn{1-\prn{v/f}^2} \Gamma^{\SM}_{h\rightarrow f}.$
Thus, the branching ratio is
\al{
\label{eq:BR h to f}
\mathcal{BR}\left(h\rightarrow f\right)=\frac{\prn{1-\prn{v/f}^2} \Gamma^{\SM}_{h\rightarrow f}}{\Gamma_h}.
}
Combining Eq.'s~\eqref{eq:mu defn} to~\eqref{eq:BR h to f}, we find
\al{
\label{eq:predicted mu}
\mu = \prn{1-\prn{v/f}^2} \left(1-\mathcal{BR}\left(h\rightarrow \inv\right)\right).
}

The most up-to-date bounds on the signal strength $\mu$ come from Ref.~\cite{ATLAS:2017ovn}. 
We don't use the global signal strength they report below Eq.~(2) because they combine many inaccurate channels to arrive at their global fit. 
Instead, we take the result for the $gg\rightarrow h_{\rm phys}$ (0-jet) from the top of Fig.~(9). 
We require our light twin benchmark to satisfy $\mu \ge 0.8$. Fig.~\ref{fig:BRhtoinv} demonstrates that our parameter space easily avoids this current bound.

\bibliographystyle{JHEP}
\bibliography{refs}

\providecommand{\href}[2]{#2}\begingroup\raggedright\begin{thebibliography}{10}

\bibitem{Murayama:2000dw}
H.~Murayama, \emph{{Supersymmetry phenomenology}},  in \emph{{Proceedings,
  Summer School in Particle Physics: Trieste, Italy, June 21-July 9, 1999}},
  pp.~296--335, 2000, \href{https://arxiv.org/abs/hep-ph/0002232}{{\ttfamily
  hep-ph/0002232}}.

\bibitem{Maiani:1979cx}
L.~Maiani, \emph{{All You Need to Know about the Higgs Boson}}, .

\bibitem{Veltman:1980mj}
M.~J.~G. Veltman, \emph{{The Infrared - Ultraviolet Connection}}, {\emph{Acta
  Phys. Polon.} {\bfseries B12} (1981) 437}.

\bibitem{Witten:1981nf}
E.~Witten, \emph{{Dynamical Breaking of Supersymmetry}},
  \href{https://doi.org/10.1016/0550-3213(81)90006-7}{\emph{Nucl. Phys.}
  {\bfseries B188} (1981) 513}.

\bibitem{Kaul:1981wp}
R.~K. Kaul, \emph{{Gauge Hierarchy in a Supersymmetric Model}},
  \href{https://doi.org/10.1016/0370-2693(82)90453-1}{\emph{Phys. Lett.}
  {\bfseries 109B} (1982) 19}.

\bibitem{Kaplan:1983fs}
D.~B. Kaplan and H.~Georgi, \emph{{SU(2) x U(1) Breaking by Vacuum
  Misalignment}},
  \href{https://doi.org/10.1016/0370-2693(84)91177-8}{\emph{Phys. Lett.}
  {\bfseries 136B} (1984) 183}.

\bibitem{Kaplan:1983sm}
D.~B. Kaplan, H.~Georgi and S.~Dimopoulos, \emph{{Composite Higgs Scalars}},
  \href{https://doi.org/10.1016/0370-2693(84)91178-X}{\emph{Phys. Lett.}
  {\bfseries 136B} (1984) 187}.

\bibitem{Aaboud:2018kya}
{\scshape ATLAS} collaboration, \emph{{Search for top squarks decaying to tau
  sleptons in $pp$ collisions at $\sqrt{s}= 13$ TeV with the ATLAS detector}},
  \href{https://doi.org/10.1103/PhysRevD.98.032008}{\emph{Phys. Rev.}
  {\bfseries D98} (2018) 032008}
  [\href{https://arxiv.org/abs/1803.10178}{{\ttfamily 1803.10178}}].

\bibitem{Sirunyan:2018vjp}
{\scshape CMS} collaboration, \emph{{Search for natural and split supersymmetry
  in proton-proton collisions at $ \sqrt{s}=13 $ TeV in final states with jets
  and missing transverse momentum}},
  \href{https://doi.org/10.1007/JHEP05(2018)025}{\emph{JHEP} {\bfseries 05}
  (2018) 025} [\href{https://arxiv.org/abs/1802.02110}{{\ttfamily
  1802.02110}}].

\bibitem{Aad:2016shx}
{\scshape ATLAS} collaboration, \emph{{Search for single production of a
  vector-like quark via a heavy gluon in the $4b$ final state with the ATLAS
  detector in $pp$ collisions at $\sqrt{s} = 8$ TeV}},
  \href{https://doi.org/10.1016/j.physletb.2016.04.061}{\emph{Phys. Lett.}
  {\bfseries B758} (2016) 249}
  [\href{https://arxiv.org/abs/1602.06034}{{\ttfamily 1602.06034}}].

\bibitem{Sirunyan:2018omb}
{\scshape CMS} collaboration, \emph{{Search for vector-like T and B quark pairs
  in final states with leptons at $\sqrt{s} =$ 13 TeV}},
  \href{https://doi.org/10.1007/JHEP08(2018)177}{\emph{JHEP} {\bfseries 08}
  (2018) 177} [\href{https://arxiv.org/abs/1805.04758}{{\ttfamily
  1805.04758}}].

\bibitem{Chacko:2005pe}
Z.~Chacko, H.-S. Goh and R.~Harnik, \emph{{The Twin Higgs: Natural electroweak
  breaking from mirror symmetry}},
  \href{https://doi.org/10.1103/PhysRevLett.96.231802}{\emph{Phys. Rev. Lett.}
  {\bfseries 96} (2006) 231802}
  [\href{https://arxiv.org/abs/hep-ph/0506256}{{\ttfamily hep-ph/0506256}}].

\bibitem{Falkowski:2006qq}
A.~Falkowski, S.~Pokorski and M.~Schmaltz, \emph{{Twin SUSY}},
  \href{https://doi.org/10.1103/PhysRevD.74.035003}{\emph{Phys. Rev.}
  {\bfseries D74} (2006) 035003}
  [\href{https://arxiv.org/abs/hep-ph/0604066}{{\ttfamily hep-ph/0604066}}].

\bibitem{Chang:2006ra}
S.~Chang, L.~J. Hall and N.~Weiner, \emph{{A Supersymmetric twin Higgs}},
  \href{https://doi.org/10.1103/PhysRevD.75.035009}{\emph{Phys. Rev.}
  {\bfseries D75} (2007) 035009}
  [\href{https://arxiv.org/abs/hep-ph/0604076}{{\ttfamily hep-ph/0604076}}].

\bibitem{Craig:2013fga}
N.~Craig and K.~Howe, \emph{{Doubling down on naturalness with a supersymmetric
  twin Higgs}}, \href{https://doi.org/10.1007/JHEP03(2014)140}{\emph{JHEP}
  {\bfseries 03} (2014) 140} [\href{https://arxiv.org/abs/1312.1341}{{\ttfamily
  1312.1341}}].

\bibitem{Katz:2016wtw}
A.~Katz, A.~Mariotti, S.~Pokorski, D.~Redigolo and R.~Ziegler, \emph{{SUSY
  Meets Her Twin}}, \href{https://doi.org/10.1007/JHEP01(2017)142}{\emph{JHEP}
  {\bfseries 01} (2017) 142}
  [\href{https://arxiv.org/abs/1611.08615}{{\ttfamily 1611.08615}}].

\bibitem{Badziak:2017syq}
M.~Badziak and K.~Harigaya, \emph{{Supersymmetric D-term Twin Higgs}},
  \href{https://doi.org/10.1007/JHEP06(2017)065}{\emph{JHEP} {\bfseries 06}
  (2017) 065} [\href{https://arxiv.org/abs/1703.02122}{{\ttfamily
  1703.02122}}].

\bibitem{Badziak:2017kjk}
M.~Badziak and K.~Harigaya, \emph{{Minimal Non-Abelian Supersymmetric Twin
  Higgs}}, \href{https://doi.org/10.1007/JHEP10(2017)109}{\emph{JHEP}
  {\bfseries 10} (2017) 109}
  [\href{https://arxiv.org/abs/1707.09071}{{\ttfamily 1707.09071}}].

\bibitem{Badziak:2017wxn}
M.~Badziak and K.~Harigaya, \emph{{Asymptotically Free Natural Supersymmetric
  Twin Higgs Model}},
  \href{https://doi.org/10.1103/PhysRevLett.120.211803}{\emph{Phys. Rev. Lett.}
  {\bfseries 120} (2018) 211803}
  [\href{https://arxiv.org/abs/1711.11040}{{\ttfamily 1711.11040}}].

\bibitem{Batra:2008jy}
P.~Batra and Z.~Chacko, \emph{{A Composite Twin Higgs Model}},
  \href{https://doi.org/10.1103/PhysRevD.79.095012}{\emph{Phys. Rev.}
  {\bfseries D79} (2009) 095012}
  [\href{https://arxiv.org/abs/0811.0394}{{\ttfamily 0811.0394}}].

\bibitem{Geller:2014kta}
M.~Geller and O.~Telem, \emph{{Holographic Twin Higgs Model}},
  \href{https://doi.org/10.1103/PhysRevLett.114.191801}{\emph{Phys. Rev. Lett.}
  {\bfseries 114} (2015) 191801}
  [\href{https://arxiv.org/abs/1411.2974}{{\ttfamily 1411.2974}}].

\bibitem{Barbieri:2015lqa}
R.~Barbieri, D.~Greco, R.~Rattazzi and A.~Wulzer, \emph{{The Composite Twin
  Higgs scenario}}, \href{https://doi.org/10.1007/JHEP08(2015)161}{\emph{JHEP}
  {\bfseries 08} (2015) 161}
  [\href{https://arxiv.org/abs/1501.07803}{{\ttfamily 1501.07803}}].

\bibitem{Low:2015nqa}
M.~Low, A.~Tesi and L.-T. Wang, \emph{{Twin Higgs mechanism and a composite
  Higgs boson}}, \href{https://doi.org/10.1103/PhysRevD.91.095012}{\emph{Phys.
  Rev.} {\bfseries D91} (2015) 095012}
  [\href{https://arxiv.org/abs/1501.07890}{{\ttfamily 1501.07890}}].

\bibitem{Cheng:2015buv}
H.-C. Cheng, S.~Jung, E.~Salvioni and Y.~Tsai, \emph{{Exotic Quarks in Twin
  Higgs Models}}, \href{https://doi.org/10.1007/JHEP03(2016)074}{\emph{JHEP}
  {\bfseries 03} (2016) 074}
  [\href{https://arxiv.org/abs/1512.02647}{{\ttfamily 1512.02647}}].

\bibitem{Csaki:2015gfd}
C.~Csaki, M.~Geller, O.~Telem and A.~Weiler, \emph{{The Flavor of the Composite
  Twin Higgs}}, \href{https://doi.org/10.1007/JHEP09(2016)146}{\emph{JHEP}
  {\bfseries 09} (2016) 146}
  [\href{https://arxiv.org/abs/1512.03427}{{\ttfamily 1512.03427}}].

\bibitem{Cheng:2016uqk}
H.-C. Cheng, E.~Salvioni and Y.~Tsai, \emph{{Exotic electroweak signals in the
  twin Higgs model}},
  \href{https://doi.org/10.1103/PhysRevD.95.115035}{\emph{Phys. Rev.}
  {\bfseries D95} (2017) 115035}
  [\href{https://arxiv.org/abs/1612.03176}{{\ttfamily 1612.03176}}].

\bibitem{Contino:2017moj}
R.~Contino, D.~Greco, R.~Mahbubani, R.~Rattazzi and R.~Torre, \emph{{Precision
  Tests and Fine Tuning in Twin Higgs Models}},
  \href{https://doi.org/10.1103/PhysRevD.96.095036}{\emph{Phys. Rev.}
  {\bfseries D96} (2017) 095036}
  [\href{https://arxiv.org/abs/1702.00797}{{\ttfamily 1702.00797}}].

\bibitem{deSalas:2016ztq}
P.~F. de~Salas and S.~Pastor, \emph{{Relic neutrino decoupling with flavour
  oscillations revisited}},
  \href{https://doi.org/10.1088/1475-7516/2016/07/051}{\emph{JCAP} {\bfseries
  1607} (2016) 051} [\href{https://arxiv.org/abs/1606.06986}{{\ttfamily
  1606.06986}}].

\bibitem{Mangano:2001iu}
G.~Mangano, G.~Miele, S.~Pastor and M.~Peloso, \emph{{A Precision calculation
  of the effective number of cosmological neutrinos}},
  \href{https://doi.org/10.1016/S0370-2693(02)01622-2}{\emph{Phys. Lett.}
  {\bfseries B534} (2002) 8}
  [\href{https://arxiv.org/abs/astro-ph/0111408}{{\ttfamily
  astro-ph/0111408}}].

\bibitem{Chacko:2016hvu}
Z.~Chacko, N.~Craig, P.~J. Fox and R.~Harnik, \emph{{Cosmology in Mirror Twin
  Higgs and Neutrino Masses}},
  \href{https://doi.org/10.1007/JHEP07(2017)023}{\emph{JHEP} {\bfseries 07}
  (2017) 023} [\href{https://arxiv.org/abs/1611.07975}{{\ttfamily
  1611.07975}}].

\bibitem{Cyburt:2015mya}
R.~H. Cyburt, B.~D. Fields, K.~A. Olive and T.-H. Yeh, \emph{{Big Bang
  Nucleosynthesis: 2015}},
  \href{https://doi.org/10.1103/RevModPhys.88.015004}{\emph{Rev. Mod. Phys.}
  {\bfseries 88} (2016) 015004}
  [\href{https://arxiv.org/abs/1505.01076}{{\ttfamily 1505.01076}}].

\bibitem{planck2018}
{\scshape Planck} collaboration, \emph{{Planck 2018 results. VI. Cosmological
  parameters}},  \href{https://arxiv.org/abs/1807.06209}{{\ttfamily
  1807.06209}}.

\bibitem{Craig:2015pha}
N.~Craig, A.~Katz, M.~Strassler and R.~Sundrum, \emph{{Naturalness in the Dark
  at the LHC}}, \href{https://doi.org/10.1007/JHEP07(2015)105}{\emph{JHEP}
  {\bfseries 07} (2015) 105}
  [\href{https://arxiv.org/abs/1501.05310}{{\ttfamily 1501.05310}}].

\bibitem{Craig:2015xla}
N.~Craig and A.~Katz, \emph{{The Fraternal WIMP Miracle}},
  \href{https://doi.org/10.1088/1475-7516/2015/10/054}{\emph{JCAP} {\bfseries
  1510} (2015) 054} [\href{https://arxiv.org/abs/1505.07113}{{\ttfamily
  1505.07113}}].

\bibitem{Barbieri:2016zxn}
R.~Barbieri, L.~J. Hall and K.~Harigaya, \emph{{Minimal Mirror Twin Higgs}},
  \href{https://doi.org/10.1007/JHEP11(2016)172}{\emph{JHEP} {\bfseries 11}
  (2016) 172} [\href{https://arxiv.org/abs/1609.05589}{{\ttfamily
  1609.05589}}].

\bibitem{Csaki:2017spo}
C.~Csaki, E.~Kuflik and S.~Lombardo, \emph{{Viable Twin Cosmology from Neutrino
  Mixing}}, \href{https://doi.org/10.1103/PhysRevD.96.055013}{\emph{Phys. Rev.}
  {\bfseries D96} (2017) 055013}
  [\href{https://arxiv.org/abs/1703.06884}{{\ttfamily 1703.06884}}].

\bibitem{Batell:2019ptb}
B.~Batell and C.~B. Verhaaren, \emph{{Breaking Mirror Twin Hypercharge}},
  \href{https://arxiv.org/abs/1904.10468}{{\ttfamily 1904.10468}}.

\bibitem{Liu:2019ixm}
D.~Liu and N.~Weiner, \emph{{A Portalino to the Twin Sector}},
  \href{https://arxiv.org/abs/1905.00861}{{\ttfamily 1905.00861}}.

\bibitem{Hochberg:2018vdo}
Y.~Hochberg, E.~Kuflik and H.~Murayama, \emph{{Twin Higgs model with strongly
  interacting massive particle dark matter}},
  \href{https://doi.org/10.1103/PhysRevD.99.015005}{\emph{Phys. Rev.}
  {\bfseries D99} (2019) 015005}
  [\href{https://arxiv.org/abs/1805.09345}{{\ttfamily 1805.09345}}].

\bibitem{Cheng:2018vaj}
H.-C. Cheng, L.~Li and R.~Zheng, \emph{{Coscattering/Coannihilation Dark Matter
  in a Fraternal Twin Higgs Model}},
  \href{https://doi.org/10.1007/JHEP09(2018)098}{\emph{JHEP} {\bfseries 09}
  (2018) 098} [\href{https://arxiv.org/abs/1805.12139}{{\ttfamily
  1805.12139}}].

\bibitem{Craig:2016lyx}
N.~Craig, S.~Koren and T.~Trott, \emph{{Cosmological Signals of a Mirror Twin
  Higgs}}, \href{https://doi.org/10.1007/JHEP05(2017)038}{\emph{JHEP}
  {\bfseries 05} (2017) 038}
  [\href{https://arxiv.org/abs/1611.07977}{{\ttfamily 1611.07977}}].

\bibitem{Craig:2016kue}
N.~Craig, S.~Knapen, P.~Longhi and M.~Strassler, \emph{{The Vector-like Twin
  Higgs}}, \href{https://doi.org/10.1007/JHEP07(2016)002}{\emph{JHEP}
  {\bfseries 07} (2016) 002}
  [\href{https://arxiv.org/abs/1601.07181}{{\ttfamily 1601.07181}}].

\bibitem{Koren:2019iuv}
S.~Koren and R.~McGehee, \emph{{Freezing-in twin dark matter}},
  \href{https://doi.org/10.1103/PhysRevD.101.055024}{\emph{Phys. Rev. D}
  {\bfseries 101} (2020) 055024}
  [\href{https://arxiv.org/abs/1908.03559}{{\ttfamily 1908.03559}}].

\bibitem{Chacko:2018vss}
Z.~Chacko, D.~Curtin, M.~Geller and Y.~Tsai, \emph{{Cosmological Signatures of
  a Mirror Twin Higgs}},
  \href{https://doi.org/10.1007/JHEP09(2018)163}{\emph{JHEP} {\bfseries 09}
  (2018) 163} [\href{https://arxiv.org/abs/1803.03263}{{\ttfamily
  1803.03263}}].

\bibitem{Barbieri:2017opf}
R.~Barbieri, L.~J. Hall and K.~Harigaya, \emph{{Effective Theory of Flavor for
  Minimal Mirror Twin Higgs}},
  \href{https://doi.org/10.1007/JHEP10(2017)015}{\emph{JHEP} {\bfseries 10}
  (2017) 015} [\href{https://arxiv.org/abs/1706.05548}{{\ttfamily
  1706.05548}}].

\bibitem{Froggatt:1978nt}
C.~D. Froggatt and H.~B. Nielsen, \emph{{Hierarchy of Quark Masses, Cabibbo
  Angles and CP Violation}},
  \href{https://doi.org/10.1016/0550-3213(79)90316-X}{\emph{Nucl. Phys.}
  {\bfseries B147} (1979) 277}.

\bibitem{Yanagida:1979as}
T.~Yanagida, \emph{{Horizontal gauge symmetry and masses of neutrinos}},
  {\emph{Conf. Proc.} {\bfseries C7902131} (1979) 95}.

\bibitem{GellMann:1980vs}
M.~Gell-Mann, P.~Ramond and R.~Slansky, \emph{{Complex Spinors and Unified
  Theories}}, {\emph{Conf. Proc.} {\bfseries C790927} (1979) 315}
  [\href{https://arxiv.org/abs/1306.4669}{{\ttfamily 1306.4669}}].

\bibitem{Minkowski:1977sc}
P.~Minkowski, \emph{{$\mu \to e\gamma$ at a Rate of One Out of $10^{9}$ Muon
  Decays?}}, \href{https://doi.org/10.1016/0370-2693(77)90435-X}{\emph{Phys.
  Lett.} {\bfseries 67B} (1977) 421}.

\bibitem{Mohapatra:1979ia}
R.~N. Mohapatra and G.~Senjanovic, \emph{{Neutrino Mass and Spontaneous Parity
  Nonconservation}},
  \href{https://doi.org/10.1103/PhysRevLett.44.912}{\emph{Phys. Rev. Lett.}
  {\bfseries 44} (1980) 912}.

\bibitem{Fukugita:1986hr}
M.~Fukugita and T.~Yanagida, \emph{{Baryogenesis Without Grand Unification}},
  \href{https://doi.org/10.1016/0370-2693(86)91126-3}{\emph{Phys. Lett.}
  {\bfseries B174} (1986) 45}.

\bibitem{Giudice:2003jh}
G.~F. Giudice, A.~Notari, M.~Raidal, A.~Riotto and A.~Strumia, \emph{{Towards a
  complete theory of thermal leptogenesis in the SM and MSSM}},
  \href{https://doi.org/10.1016/j.nuclphysb.2004.02.019}{\emph{Nucl. Phys.}
  {\bfseries B685} (2004) 89}
  [\href{https://arxiv.org/abs/hep-ph/0310123}{{\ttfamily hep-ph/0310123}}].

\bibitem{Buchmuller:2004nz}
W.~Buchmuller, P.~Di~Bari and M.~Plumacher, \emph{{Leptogenesis for
  pedestrians}}, \href{https://doi.org/10.1016/j.aop.2004.02.003}{\emph{Annals
  Phys.} {\bfseries 315} (2005) 305}
  [\href{https://arxiv.org/abs/hep-ph/0401240}{{\ttfamily hep-ph/0401240}}].

\bibitem{Khachatryan:2016vau}
{\scshape ATLAS, CMS} collaboration, \emph{{Measurements of the Higgs boson
  production and decay rates and constraints on its couplings from a combined
  ATLAS and CMS analysis of the LHC pp collision data at $ \sqrt{s}=7 $ and 8
  TeV}}, \href{https://doi.org/10.1007/JHEP08(2016)045}{\emph{JHEP} {\bfseries
  08} (2016) 045} [\href{https://arxiv.org/abs/1606.02266}{{\ttfamily
  1606.02266}}].

\bibitem{Chang:2016ntp}
J.~H. Chang, R.~Essig and S.~D. McDermott, \emph{{Revisiting Supernova 1987A
  Constraints on Dark Photons}},
  \href{https://doi.org/10.1007/JHEP01(2017)107}{\emph{JHEP} {\bfseries 01}
  (2017) 107} [\href{https://arxiv.org/abs/1611.03864}{{\ttfamily
  1611.03864}}].

\bibitem{DeRocco:2019njg}
W.~DeRocco, P.~W. Graham, D.~Kasen, G.~Marques-Tavares and S.~Rajendran,
  \emph{{Observable signatures of dark photons from supernovae}},
  \href{https://doi.org/10.1007/JHEP02(2019)171}{\emph{JHEP} {\bfseries 02}
  (2019) 171} [\href{https://arxiv.org/abs/1901.08596}{{\ttfamily
  1901.08596}}].

\bibitem{Sung:2019xie}
A.~Sung, H.~Tu and M.-R. Wu, \emph{{New constraint from supernova explosions on
  light particles beyond the Standard Model}},
  \href{https://arxiv.org/abs/1903.07923}{{\ttfamily 1903.07923}}.

\bibitem{Alexander:2016aln}
J.~Alexander et~al., \emph{{Dark Sectors 2016 Workshop: Community Report}},
  2016, \href{https://arxiv.org/abs/1608.08632}{{\ttfamily 1608.08632}},
  \href{http://lss.fnal.gov/archive/2016/conf/fermilab-conf-16-421.pdf}{http://lss.fnal.gov/archive/2016/conf/fermilab-conf-16-421.pdf}.

\bibitem{Parker:2018vye}
R.~H. Parker, C.~Yu, W.~Zhong, B.~Estey and H.~M{\"u}ller, \emph{{Measurement
  of the fine-structure constant as a test of the Standard Model}},
  \href{https://doi.org/10.1126/science.aap7706}{\emph{Science} {\bfseries 360}
  (2018) 191} [\href{https://arxiv.org/abs/1812.04130}{{\ttfamily
  1812.04130}}].

\bibitem{Berlin:2018bsc}
A.~Berlin, N.~Blinov, G.~Krnjaic, P.~Schuster and N.~Toro, \emph{{Dark Matter,
  Millicharges, Axion and Scalar Particles, Gauge Bosons, and Other New Physics
  with LDMX}},  \href{https://arxiv.org/abs/1807.01730}{{\ttfamily
  1807.01730}}.

\bibitem{Echenard:2014lma}
B.~Echenard, R.~Essig and Y.-M. Zhong, \emph{{Projections for Dark Photon
  Searches at Mu3e}},
  \href{https://doi.org/10.1007/JHEP01(2015)113}{\emph{JHEP} {\bfseries 01}
  (2015) 113} [\href{https://arxiv.org/abs/1411.1770}{{\ttfamily 1411.1770}}].

\bibitem{Berlin:2018pwi}
A.~Berlin, S.~Gori, P.~Schuster and N.~Toro, \emph{{Dark Sectors at the
  Fermilab SeaQuest Experiment}},
  \href{https://doi.org/10.1103/PhysRevD.98.035011}{\emph{Phys. Rev.}
  {\bfseries D98} (2018) 035011}
  [\href{https://arxiv.org/abs/1804.00661}{{\ttfamily 1804.00661}}].

\bibitem{Celentano:2014wya}
{\scshape HPS} collaboration, \emph{{The Heavy Photon Search experiment at
  Jefferson Laboratory}},
  \href{https://doi.org/10.1088/1742-6596/556/1/012064}{\emph{J. Phys. Conf.
  Ser.} {\bfseries 556} (2014) 012064}
  [\href{https://arxiv.org/abs/1505.02025}{{\ttfamily 1505.02025}}].

\bibitem{Alekhin:2015byh}
S.~Alekhin et~al., \emph{{A facility to Search for Hidden Particles at the CERN
  SPS: the SHiP physics case}},
  \href{https://doi.org/10.1088/0034-4885/79/12/124201}{\emph{Rept. Prog.
  Phys.} {\bfseries 79} (2016) 124201}
  [\href{https://arxiv.org/abs/1504.04855}{{\ttfamily 1504.04855}}].

\bibitem{Ariga:2018uku}
{\scshape FASER} collaboration, \emph{{FASER's Physics Reach for Long-Lived
  Particles}},  \href{https://arxiv.org/abs/1811.12522}{{\ttfamily
  1811.12522}}.

\bibitem{Lanfranchi:2017wzl}
{\scshape NA62} collaboration, \emph{{Search for Hidden Sector particles at
  NA62}}, \href{https://doi.org/10.22323/1.314.0301}{\emph{PoS} {\bfseries
  EPS-HEP2017} (2017) 301}.

\bibitem{Tanabashi:2018oca}
{\scshape Particle Data Group} collaboration, \emph{{Review of Particle
  Physics}}, \href{https://doi.org/10.1103/PhysRevD.98.030001}{\emph{Phys.
  Rev.} {\bfseries D98} (2018) 030001}.

\bibitem{Bambade:2019fyw}
P.~Bambade et~al., \emph{{The International Linear Collider: A Global
  Project}},  \href{https://arxiv.org/abs/1903.01629}{{\ttfamily 1903.01629}}.

\bibitem{Henning:2015alf}
B.~Henning, X.~Lu, T.~Melia and H.~Murayama, \emph{{2, 84, 30, 993, 560, 15456,
  11962, 261485, ...: Higher dimension operators in the SM EFT}},
  \href{https://doi.org/10.1007/JHEP08(2017)016}{\emph{JHEP} {\bfseries 08}
  (2017) 016} [\href{https://arxiv.org/abs/1512.03433}{{\ttfamily
  1512.03433}}].

\bibitem{Novikov:1977dq}
V.~A. Novikov, L.~B. Okun, M.~A. Shifman, A.~I. Vainshtein, M.~B. Voloshin and
  V.~I. Zakharov, \emph{{Charmonium and Gluons: Basic Experimental Facts and
  Theoretical Introduction}},
  \href{https://doi.org/10.1016/0370-1573(78)90120-5}{\emph{Phys. Rept.}
  {\bfseries 41} (1978) 1}.

\bibitem{Mangano:2006ar}
G.~Mangano, G.~Miele, S.~Pastor, T.~Pinto, O.~Pisanti and P.~D. Serpico,
  \emph{{Effects of non-standard neutrino-electron interactions on relic
  neutrino decoupling}},
  \href{https://doi.org/10.1016/j.nuclphysb.2006.09.002}{\emph{Nucl. Phys.}
  {\bfseries B756} (2006) 100}
  [\href{https://arxiv.org/abs/hep-ph/0607267}{{\ttfamily hep-ph/0607267}}].

\bibitem{Aver:2015iza}
E.~Aver, K.~A. Olive and E.~D. Skillman, \emph{{The effects of He I
  $\lambda$10830 on helium abundance determinations}},
  \href{https://doi.org/10.1088/1475-7516/2015/07/011}{\emph{JCAP} {\bfseries
  1507} (2015) 011} [\href{https://arxiv.org/abs/1503.08146}{{\ttfamily
  1503.08146}}].

\bibitem{Galvez:2016sza}
R.~Galvez and R.~J. Scherrer, \emph{{Cosmology with Independently Varying
  Neutrino Temperature and Number}},
  \href{https://doi.org/10.1103/PhysRevD.95.063507}{\emph{Phys. Rev.}
  {\bfseries D95} (2017) 063507}
  [\href{https://arxiv.org/abs/1609.06351}{{\ttfamily 1609.06351}}].

\bibitem{Arbey:2011nf}
A.~Arbey, \emph{{AlterBBN: A program for calculating the BBN abundances of the
  elements in alternative cosmologies}},
  \href{https://doi.org/10.1016/j.cpc.2012.03.018}{\emph{Comput. Phys. Commun.}
  {\bfseries 183} (2012) 1822}
  [\href{https://arxiv.org/abs/1106.1363}{{\ttfamily 1106.1363}}].

\bibitem{Arbey:2018zfh}
A.~Arbey, J.~Auffinger, K.~P. Hickerson and E.~S. Jenssen, \emph{{AlterBBN v2:
  A public code for calculating Big-Bang nucleosynthesis constraints in
  alternative cosmologies}},
  \href{https://arxiv.org/abs/1806.11095}{{\ttfamily 1806.11095}}.

\bibitem{Benson:2014qhw}
{\scshape SPT-3G} collaboration, \emph{{SPT-3G: A Next-Generation Cosmic
  Microwave Background Polarization Experiment on the South Pole Telescope}},
  \href{https://doi.org/10.1117/12.2057305}{\emph{Proc. SPIE Int. Soc. Opt.
  Eng.} {\bfseries 9153} (2014) 91531P}
  [\href{https://arxiv.org/abs/1407.2973}{{\ttfamily 1407.2973}}].

\bibitem{Louis:2016ahn}
{\scshape ACTPol} collaboration, \emph{{The Atacama Cosmology Telescope:
  Two-Season ACTPol Spectra and Parameters}},
  \href{https://doi.org/10.1088/1475-7516/2017/06/031}{\emph{JCAP} {\bfseries
  1706} (2017) 031} [\href{https://arxiv.org/abs/1610.02360}{{\ttfamily
  1610.02360}}].

\bibitem{Suzuki:2015zzg}
{\scshape POLARBEAR} collaboration, \emph{{The POLARBEAR-2 and the Simons Array
  Experiment}}, \href{https://doi.org/10.1007/s10909-015-1425-4}{\emph{J. Low.
  Temp. Phys.} {\bfseries 184} (2016) 805}
  [\href{https://arxiv.org/abs/1512.07299}{{\ttfamily 1512.07299}}].

\bibitem{Grayson:2016smb}
{\scshape BICEP3} collaboration, \emph{{BICEP3 performance overview and planned
  Keck Array upgrade}}, \href{https://doi.org/10.1117/12.2233894}{\emph{Proc.
  SPIE Int. Soc. Opt. Eng.} {\bfseries 9914} (2016) 99140S}
  [\href{https://arxiv.org/abs/1607.04668}{{\ttfamily 1607.04668}}].

\bibitem{CMBS4}
{\scshape CMB-S4} collaboration, \emph{{CMB-S4 Science Book, First Edition}},
  \href{https://arxiv.org/abs/1610.02743}{{\ttfamily 1610.02743}}.

\bibitem{Denner:2011mq}
A.~Denner, S.~Heinemeyer, I.~Puljak, D.~Rebuzzi and M.~Spira, \emph{{Standard
  Model Higgs-Boson Branching Ratios with Uncertainties}},
  \href{https://doi.org/10.1140/epjc/s10052-011-1753-8}{\emph{Eur. Phys. J.}
  {\bfseries C71} (2011) 1753}
  [\href{https://arxiv.org/abs/1107.5909}{{\ttfamily 1107.5909}}].

\bibitem{ATLAS:2017ovn}
{\scshape ATLAS} collaboration, \emph{{Combined measurements of Higgs boson
  production and decay in the $H\rightarrow ZZ^\ast\rightarrow4\ell$ and
  $H\rightarrow\gamma\gamma$ channels using $\sqrt{s}=$ 13 TeV pp collision
  data collected with the ATLAS experiment}}, .

\end{thebibliography}\endgroup
\end{document}